\documentclass[useAMS,usenatbib,a4paper]{mn2e}
\usepackage{aas_macros}
\usepackage{graphicx}
\usepackage{mathptmx}
\usepackage{amsmath}
\usepackage{color}
\usepackage{lineno}

\newcommand{\<}{\begin{eqnarray}}
\renewcommand{\>}{\end{eqnarray}} 
\renewcommand{\bar}{\overline}
\renewcommand{\tilde}{\widetilde}
\renewcommand{\hat}{\widehat}
\newcommand{\Msun}{\ensuremath{\rmn{M}_\odot}}
\renewcommand{\(}{\left(}
\renewcommand{\)}{\right)}
\newcommand{\e}{\text{e}}
\renewcommand{\d}{\mathrm{d}} 

\newlength{\halfcolumn}
\newlength{\fullcolumn}
\newlength{\fullcolumnspace}
\title[Accretion and the IMF]{
The relation between accretion rates and the initial mass function in hydrodynamical simulations of star formation
}
\author[Th. Maschberger, I.A. Bonnell, C.J. Clarke and E. Moraux]
{Th. Maschberger$^{1}$\thanks{email: thomas.maschberger@obs.ujf-grenoble.fr},
I.A. Bonnell$^2$,
C.J. Clarke$^3$ and
E. Moraux $^1$\\
\small \it 
$ö1$ UJF-Grenoble 1/CNRS-INSU, Institut de Plan{\'e}tologie et d{'}Astrophysique de Grenoble (IPAG) UMR 5274, Grenoble, F-38041, France\\
$^2$ Scottish Universities Physics Alliance (SUPA), School of Physics and Astronomy, University of St. Andrews, North Haugh, St. Andrews, Fife KY16 9SS\\
$^3$ Institute of Astronomy, Madingley Road, Cambridge CB3 0HA\\
}
\date{}
\pagerange{\pageref{firstpage}--\pageref{lastpage}} \pubyear{201X}

\begin{document}
\maketitle
\label{firstpage}

\begin{abstract}
We analyse a hydrodynamical simulation of star formation.
Sink particles in the simulations which represent stars show episodic growth, which is presumably accretion from a core that can be regularly replenished in response to the fluctuating conditions in the local environment.
The accretion rates follow $\dot{m} \propto m^{2/3}$, as expected from accretion in a gas-dominated potential, but with substantial variations over-laid on this.
The growth times follow an exponential distribution which is tapered at long times due to the finite length of the simulation.
The initial collapse masses have an approximately lognormal distribution with already an onset of a power-law at large masses.
The sink particle mass function can be reproduced with a non-linear stochastic process, with fluctuating accretion rates $\propto m^{2/3}$,  a distribution of seed masses and a distribution of growth times.
All three factors contribute equally to the form of the final sink mass function.
We find that the upper power law tail of the IMF is unrelated to Bondi--Hoyle accretion.
\end{abstract}


\section{Introduction}

The origin of the stellar initial mass function (IMF) is a key question for a theory of star formation.
Several ideas have been proposed to explain the stellar initial mass function, for example fragmentation, competitive accretion, a distribution of growth times, or, more statistically, space filling and gravoturbulent fragmentation.
They succeed in explaining one or more properties of the IMF, such as its lognormal-like shape in the low-mass regime, the power-law behaviour at high masses (in particular the Salpeter exponent), its peak and its width.
It is the purpose of this article to investigate which of the ideas mentioned above contribute to the development of the sink particle mass function {\it in a hydrodynamical simulation of star formation} (`sink particles' are henceforth termed `sinks' throughout the paper).
We aim in the process to shed some light on the origin of the {\it observed} IMF.

Fragmentation is one of the first processes proposed for star formation, going back to 
\citet{Hoyle-1953a} and extended by a random component by  \citet{Marcus-1968a}, \citet{Larson-1973a}, \citet{ElmegreenMathieu-1983a} and \citet{Zinnecker-1984a}.
This random fragmentation, repeatedly splitting a fragment, is essentially a linear stochastic process, first described by \citet{Kolmogorov-1941a}, that leads to a lognormal distribution.
The model of \citet{Marcus-1968a} predicts also the total number of fragments in addition to their mass distribution.

Another principal concept of star formation is stellar accretion, either $\dot{m} \propto m^2$ (Bondi---Hoyle) in stellar dominated potentials  \citep{Zinnecker-1982a,BonnellBateClarke-2001a,BonnellClarkeBate-2001a}
 or $\dot{m} \propto m^{2/3}$ in gas-dominated potentials  \citep{BonnellBateClarke-2001a,BonnellClarkeBate-2001a}.
This leads to a power-law behaviour of the mass function by spreading the initial seed distribution.
In this model the power law exponent of the accretion rate-sink mass dependence is critical in determining the slope of the upper power law of the IMF and  \citet{Zinnecker-1982a} used this to relate the observed Salpeter exponent to Bondi--Hoyle accretion. 
In such models the seed distribution is the random element, both the accretion rates and growth times are not assumed to have a distribution.

A third principal concept of star formation is the distribution of growth times. 
Accretion has to stop at some point, which is likely to be a random variable.
Typically an exponential distribution of growth times is assumed \citep[e.g. ][]{Myers-2000a,Myers-2009a,ReipurthClarke-2001a,BasuJones-2004a,BateBonnell-2005a}, which implies that the probability for `killing' growth is constant in time for each star. 
The distribution of growth times leads to a distribution in mass and affects the high-mass end of the mass function.

Gravoturbulent fragmentation, with its main theories of \citet{PadoanNordlundJones-1997a, PadoanNordlund-2002a} and  \citet{HennebelleChabrier-2008a, HennebelleChabrier-2009a,HennebelleChabrier-2013a} is based on counting Jeans-unstable regions in a gas distribution that has lognormal density fluctuations superimposed by turbulence.
This is not so much related to the random splitting flavour of fragmentation mentioned above, but more related to the process of random or subdivision of a volume \citep{AuluckKothari-1954a}, which has also served in several variations as explanation for the IMF \citep[e.g. ][]{AuluckKothari-1965a,Kiang-1966a,Richtler-1994a}.
The theories of gravoturbulent fragmentation produce a mass function with a lognormal body and a power-law tail.

Several authors have attempted to combine one or more aspects, for example:
\citet{BasuJones-2004a} combine a lognormal distribution of seed masses with (deterministic) growth $\dot{m} \propto m$ or $\dot{m} \propto m^{2/3}$ and an exponential distribution of growth times.
\citet{BateBonnell-2005a} combine constant growth ($\dot{m} = \text{const}$) with a lognormal distribution of accretion rates and an exponential distribution of growth times.
(This is mathematically similar to random fragmentation models (apart from a change of sign of the quantity added); in the fragmentation models discussed above a uniform or Gaussian instead of a lognormal distribution is typically used).
\cite{Myers-2011a,Myers-2012a} investigates growth following $\dot{m} = \text{const} + m^{1.2}$ with an exponential time distribution but without a distribution of seed masses.
\cite{DibShadmehriPadoan-2010a} consider deterministic growth ($\dot{m} \propto m^{0.65}$) from a seed mass distribution given by gravoturbulent fragmentation with an exponential distribution of growth times.
\cite{Maschberger-2013b} discusses non-linear stochastic processes (a combination of random fragmentation and accretion) with growth $\dot{m} \propto m^{\alpha}$ having a lognormal distribution of accretion rates (due to the lognormal distribution of turbulent density), with a distribution of seed masses and a distribution of growth times.
This is effectively a combination of all the processes discussed above, and we will use this prescription to model the sink mass function.

Numerical studies of star formation have been performed 
on core scales \citep[typically $\approx 1\ \Msun$, 10 sinks, e.g. ][]{GoodwinWhitworthWard-Thompson-2004a,GoodwinWhitworthWard-Thompson-2004b,KrumholzKleinMcKee-2007a,VorobyovBasu-2006a,VorobyovBasu-2009a}
on small cloud scales
\citep[$\approx 100\ \Msun$, 100 sinks, e.g.][]{BateBonnellBromm-2003a,BateBonnell-2005a,Bate-2009b,Bate-2009c,Klessen-2001a,SchmejaKlessen-2004a,GirichidisFederrathBanerjee-2011a,GirichidisFederrathAllison-2012a,GirichidisFederrathBanerjee-2012a,SeifriedBanerjeeKlessen-2011a,SeifriedPudritzBanerjee-2012a,
OffnerKleinMcKee-2009a,OffnerKratterMatzner-2010a,HansenKleinMcKee-2012a,MyersMcKeeCunningham-2013a,
HennebelleCommerconJoos-2011a,CommerconHennebelleHenning-2011a}
 and on star cluster scales
\citep[$\approx 1000\ \Msun$, 1000 sinks, e.g.][]{BonnellBateVine-2003a,BonnellClarkBate-2008a,BonnellSmithClark-2011a,Bate-2009a,Bate-2012a,PetersBanerjeeKlessen-2010a,KrumholzKleinMcKee-2011a,KrumholzKleinMcKee-2012a,OffnerKleinMcKee-2008a,OffnerHansenKrumholz-2009a}
Although the simulations employ different physical processes (isothermal vs. barotropic vs. radiative transfer; wind feedback; magnetic fields; etc.) they usually lead to a sink mass function fairly similar to the IMF, if enough sinks are formed.

There have been some comparisons of theoretical models with simulations.
For example, \citet{SchmidtKernFederrath-2010a} compare the core distributions of their simulations with the models of \citet{PadoanNordlund-2002a} and \citet{HennebelleChabrier-2008a}, while 
\citet{Bate-2009a,Bate-2012a} compares the sink mass function with the model of \citet{BateBonnell-2005a}. 
In this work we set out to investigate the simulation by \citet{BonnellClarkBate-2008a,BonnellSmithClark-2011a} for the distribution and mass dependence of the accretion rates, the distribution of growth times and the distribution of seed masses in order to find out what the parameters are and where, if there is any, the main random component of star formation originates.

This paper is structured as follows:
In Section \ref{sec_calculation} we describe the calculation.
An analysis of episodic growth and the classification of sink histories follow in Sections \ref{sec_episodic_growth} and \ref{sec_classification}.
The distribution and mass dependence of the accretion rates is analysed in Section \ref{sec_accretion_rates}.
In Section \ref{sec_mass_fn} we discuss the location of each sink class in the mass function and the distribution of initial collapse masses.
Section \ref{sec_growth_times} contains the analysis of the distribution of growth times.
In Section \ref{sec_imf} we investigate which one of the random elements,  seed masses, accretion rates, and growth times, is likely the main contributor to the shape of the IMF.
A summary in Section \ref{sec_conclusions} concludes the article.

\section{Calculation}\label{sec_calculation}

We analyse the calculation performed by \citet{BonnellClarkBate-2008a,BonnellSmithClark-2011a}, to which we refer for further details.
The initial cloud mass is $10^4\ \Msun$, distributed over a cylinder 10 pc long and 3 pc in diameter.
There is a linear density gradient along the main axis, so at one end the cylinder  is 33 per cent more dense than average and at the other end 33 per cent less dense.
Turbulence is modelled by an initial divergence-free Gaussian random velocity field with a power spectrum $P(k) \propto k^4$.
Turbulence is not driven during the calculation.
At the start of the calculation the cloud is globally marginally unbound, but due to the density gradient bound in the upper half and unbound in the lower half.

Particle splitting \citep{KitsionasWhitworth-2002a,KitsionasWhitworth-2007a} was used to resolve fragmentation down to masses of 0.0167 \Msun\ (equivalent fo $4.5\ \times\ 10^7$ SPH particles), sufficient to resolve the formation of brown dwarfs. 
A lower resolution simulation was run initially to identify these regions in the initial conditions and the full simulation was then rerun, including the regions of higher resolution, in order to resolve the formation of all stars and brown dwarfs.
The gas follows a barotropic equation of state,
\< P &=& k \rho^\gamma \>
where
\< 
\begin{array}{r@{\ }c@{\ }l@{}lr@{\ }c@{\ }c@{\ }c@{\ }l}
\gamma &= &  0.&75; &          &        & \rho & \leq & \rho_1 \\
\gamma &= & 1.&0;  & \rho_1 & \leq & \rho & \leq & \rho_2 \\ 
\gamma &= & 1.&4;  & \rho_2 & \leq & \rho & \leq & \rho_3 \\ 
\gamma &= & 1.&0;  & \rho_3 & \leq & \rho &        & \\ 
\end{array}
\>
and 
$\rho_1 = 5.5 \times 10^{-19}\ \mathrm{g}\ \mathrm{cm}^{-3}$,
$\rho_2 = 5.5 \times 10^{-15}\ \mathrm{g}\ \mathrm{cm}^{-3}$ and 
$\rho_3 = 2 \times 10^{-13}\ \mathrm{g}\ \mathrm{cm}^{-3}$.
Star formation is modelled with sink particles \citep{BateBonnellPrice-1995a}, which are created at a critical density of $6.8\ \times\ 10^{-14} \text{g}\ \text{cm}^{-3}$.
The sink radius is 200 au and the accretion radius is 40 au,  gravitational interactions are also smoothed at 40 au.

The simulation runs for about one free-fall time or $6.6 \times 10^5\ \text{yr}$ and sinks start forming after $\approx 1/2 t_\text{ff}$.
In total 2542 sinks are formed with masses ranging from 0.017 to 30\ \Msun.
The vast majority of the sink particles forms in the bound half of the cylinder and is concentrated in only a few rich subclusters.

This calculation has been first published by \citet{BonnellClarkBate-2008a} who analysed it with respect to brown dwarf formation.
The evolution of subclusters, mass segregation on a subcluster scale and the upper end of the sink mass function (time variation of the exponent and truncation) has been investigated by \citet{MaschbergerClarkeBonnell-2010a}.
\citet{BonnellSmithClark-2011a} discussed the star formation efficiency in clustered and distributed regions.
The properties of cores that form in the simulation were analysed by \citet{SmithClarkBonnell-2009a,SmithLongmoreBonnell-2009a,
SmithGloverBonnell-2011a,SmithShettyStutz-2012a,SmithShettyBeuther-2013a}.
Global mass segregation was covered in \citet{MaschbergerClarke-2011a}.
\citet{KruijssenMaschbergerMoeckel-2012a} studied the dynamical structure of the subclusters, finding that they are close to virial equilibrium. 
The spatial and kinematic distribution of the sinks at the end of the calculation was dynamically evolved by \citet{MoeckelHollandClarke-2012a}, assuming instantaneous gas dispersal.

\section{Episodic growth}\label{sec_episodic_growth}

\begin{figure*}
\includegraphics[width=8.5cm]{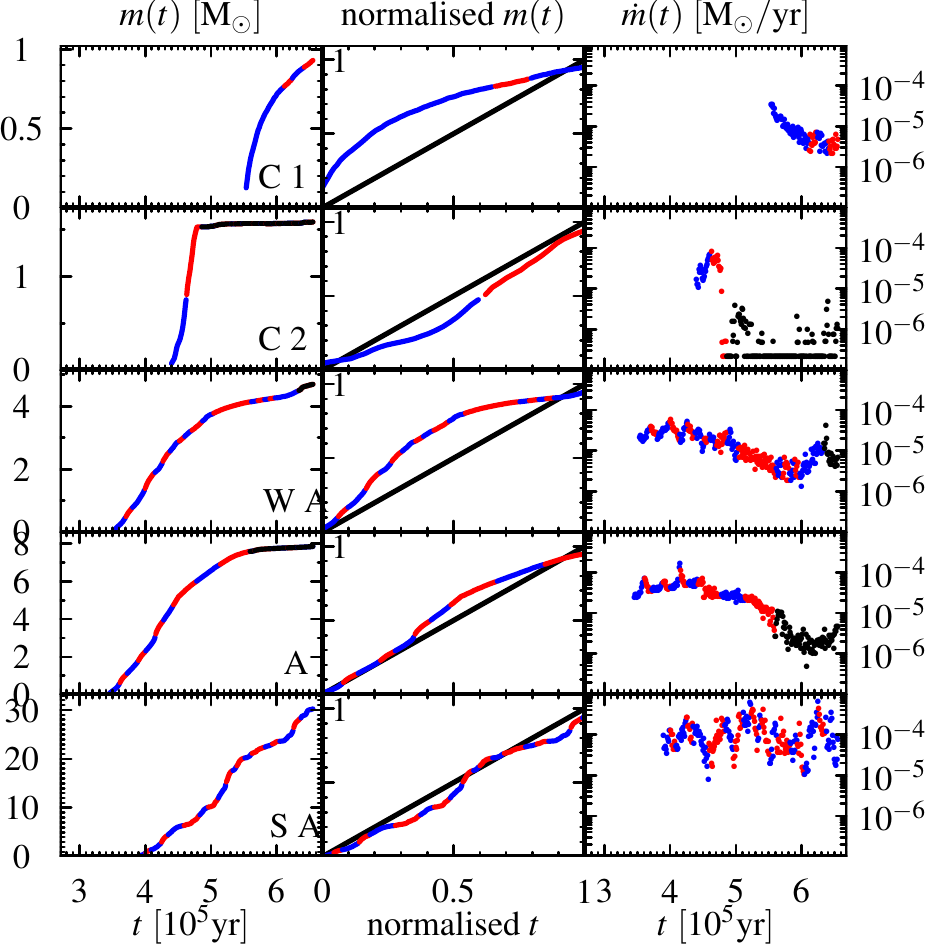}
\includegraphics[width=8.5cm]{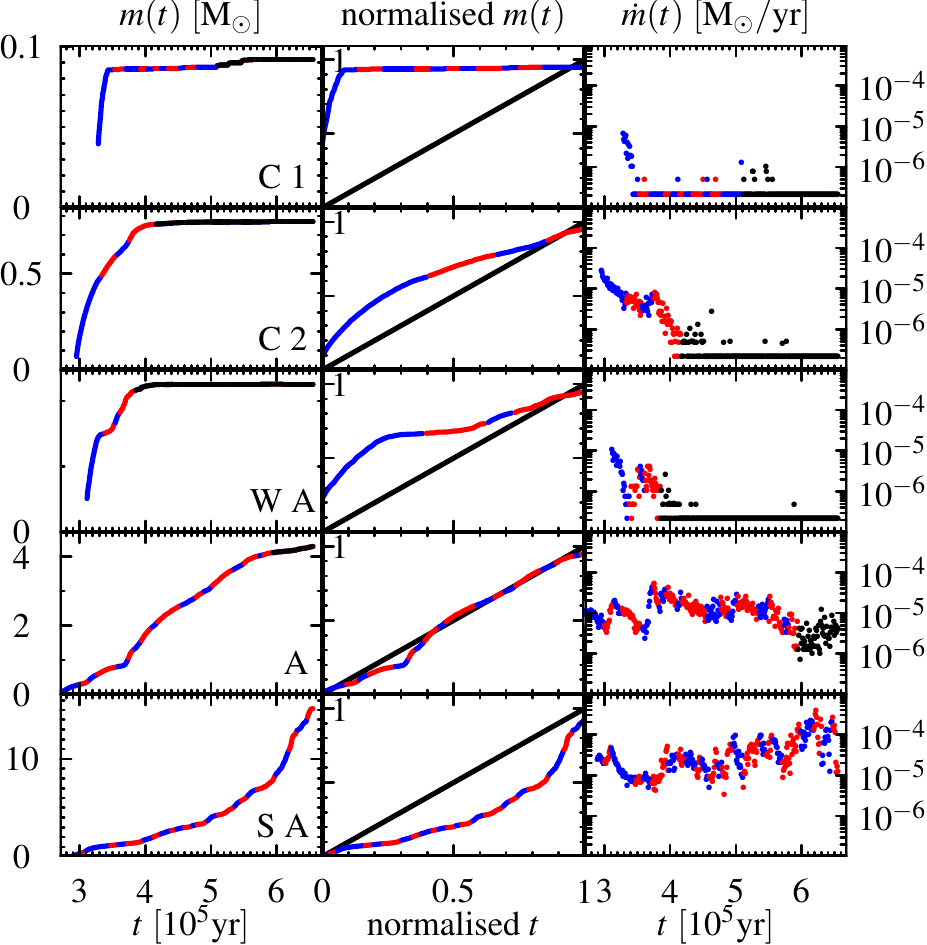}
\caption{\label{figure_example_histories}
Examples of growth histories for each of the classes (C 1= collapse in 1 episode, C 2=collapse in 2 episodes, W A=weak accretion, A=accretion, S A=strong accretion). 
The left plot shows the most massive sink in each class and the right plot the first formed sink of each class.
The left hand panels show mass as a function of time, which is normalised in the middle panels (for comparison with Fig. \ref{figure_characteristic_histories}). 
The right hand panels show the accretion rate as a function of time.
}
\end{figure*}

Fig. \ref{figure_example_histories} shows examples for the growth histories of some sinks (sampling interval 1000 yr).
The left panels give mass as a function of time, which is also shown in the middle panels, but normalised to growth time and final mass (discussed in the next section).
The right panels show the accretion rate as a function of time.
The alternating colour coding corresponds to the different episodes, whose identification is discussed in the next section.
Typically $m(t)$ in the top rows has a concave shape which becomes gradually more convex to the bottom row.
Such a behaviour of $m (t)$ is 
also seen in other simulations of star formation employing other codes and other physical processes (see e.g. fig. 2 of \citealp{PetersKlessenMac-Low-2010a}; fig. 12 of \citealp{KrumholzKleinMcKee-2011a}; fig 14 and 15 of 
\citealp{GirichidisFederrathBanerjee-2012a}; fig. 2 of \citealp{BonnellClarkeBate-2006a}).

The growth histories of the two top rows (i.e. $m(t)$ and $\dot{m} (t)$) can be understood as the collapse of an unstable  core \citep[cf. e.g.][]{FosterChevalier-1993a,WhitworthWard-Thompson-2001a}, which leads to a sharp rise of $\dot{m} (t)$ followed by an exponential-like decay.
Sink particles are created during this collapse, instantaneously collecting all gas particles fulfilling the sink creation criteria, so that only part of the collapse is traced by the sink particle mass growth.
Particularly, if the increase of $\dot{m} (t)$ is very fast the conditions for sink formation are only satisfied when $\dot{m}(t)$ is already decreasing (top row of Fig. \ref{figure_example_histories}).
This behaviour of $\dot{m} (t)$ is in agreement with the properties of the bound cores \citep{SmithGloverBonnell-2011a}.
$\dot{m}(t)$ of these lower-mass sinks is comparable to \citet{SchmejaKlessen-2004a}, \citet{GoodwinWhitworthWard-Thompson-2004a} or \citet{GirichidisFederrathBanerjee-2012a}, which are starting with a smaller gas mass.

\begin{figure}
\begin{center}
\includegraphics[width=6cm]{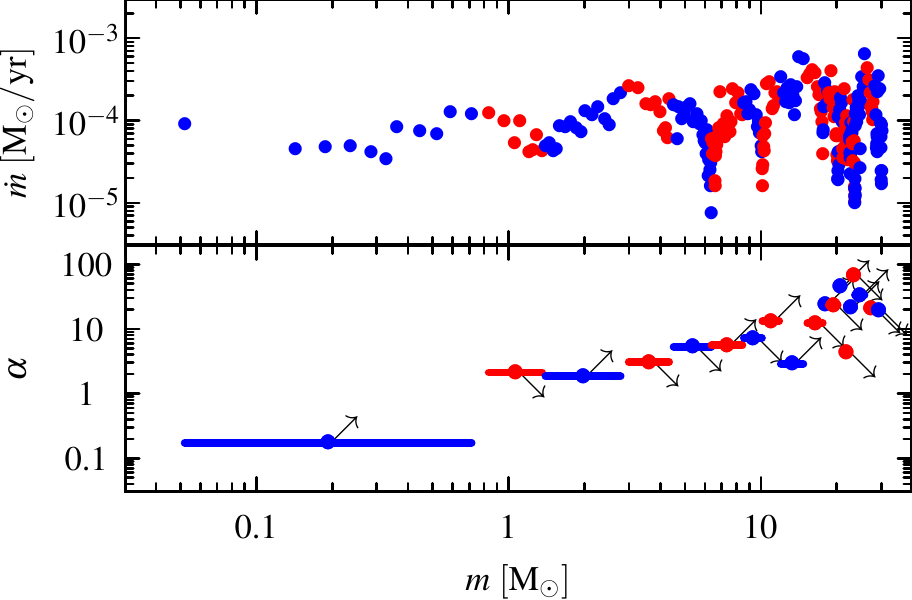}
\end{center}
\caption{\label{figure_exponents}
Top panel: accretion rate as a function of mass for the most massive sink particle.
Episodes have been colour-coded alternatingly.\newline
Bottom panel: absolute value of the fitted exponent $\alpha$ of the accretion rate $\dot{m} \propto m^{\alpha}$ for each episode.
The arrows show the sign of $\alpha$, downwards for negative $\alpha$ and upwards for positive $\alpha$.
Note the logarithmic axis for the exponent.
}
\end{figure}

In the lower panels the sinks undergo several of these accretion/collapse episodes, which leads to sometimes severe variations in $\dot{m}$, but lesser changes in the shape of $m(t)$.
This is similar to the fragmentation-induced starvation scenario of \citet{PetersBanerjeeKlessen-2010a,PetersKlessenMac-Low-2010a}.
During each episode the accretion rate  can be modelled by an exponential increase or decrease as a function of time,
\< \dot{m} (t) &=& A \e^{ b t }. \>
$A$ and $b$ are different for each episode.
This episodic growth is reflected in the plot of the accretion rates as a function of mass.
Fig. \ref{figure_exponents} (top panel) shows $\dot{m} (m)$ for the most massive sink of the simulation (also show in the bottom row of the left plot in Fig. \ref{figure_example_histories}).
Due to the episodic growth $\dot{m}$ does not depend smoothly on $m$, but shows `icicles', where $\dot{m}$ drastically decreases and $m$ is not changing much.
The simulations of \citet[][ fig. 13]{KrumholzKleinMcKee-2012a} show a similar behaviour of $\dot{m} (m)$.

The lower panel of Fig. \ref{figure_exponents} shows the absolute values for exponent of a fit  $\dot{m} \propto m^{\alpha}$ in each episode on a logarithmic scale.
Upwards arrows indicate a positive exponents (increasing $\dot{m}$) and downwards arrows a negative exponent (decreasing $\dot{m}$).
There are large variations in $\alpha$.
Although there are some episodes with $\alpha \approx 2$, generally we obtain much larger values during an episode.
It is thus hard to explain the form of the sink mass function in terms of the Bondi Hoyle accretion model proposed by \citet{Zinnecker-1982a}.

The episodic accretion that is described here is due to the repeated creation and depletion of a gas core around a sink particle {\it while the star gains most of its mass}.
This is different from the episodic accretion described in \citet{StamatellosWhitworthHubber-2011a,StamatellosWhitworthHubber-2012a} which operates in discs smaller than the sink radius of our calculation located in cores that are not replenished.
Also, the episodic accretion here has to be distinguished from bursty accretion during a T Tauri or FU Orionis phase occurring only after most of a star's mass has been assembled \citep[cf.][]{VorobyovBasu-2006a,VorobyovBasu-2009a}.

\section{Classification of the sinks}\label{sec_classification}

\begin{figure}
\begin{center}
\includegraphics[width=8.5cm]{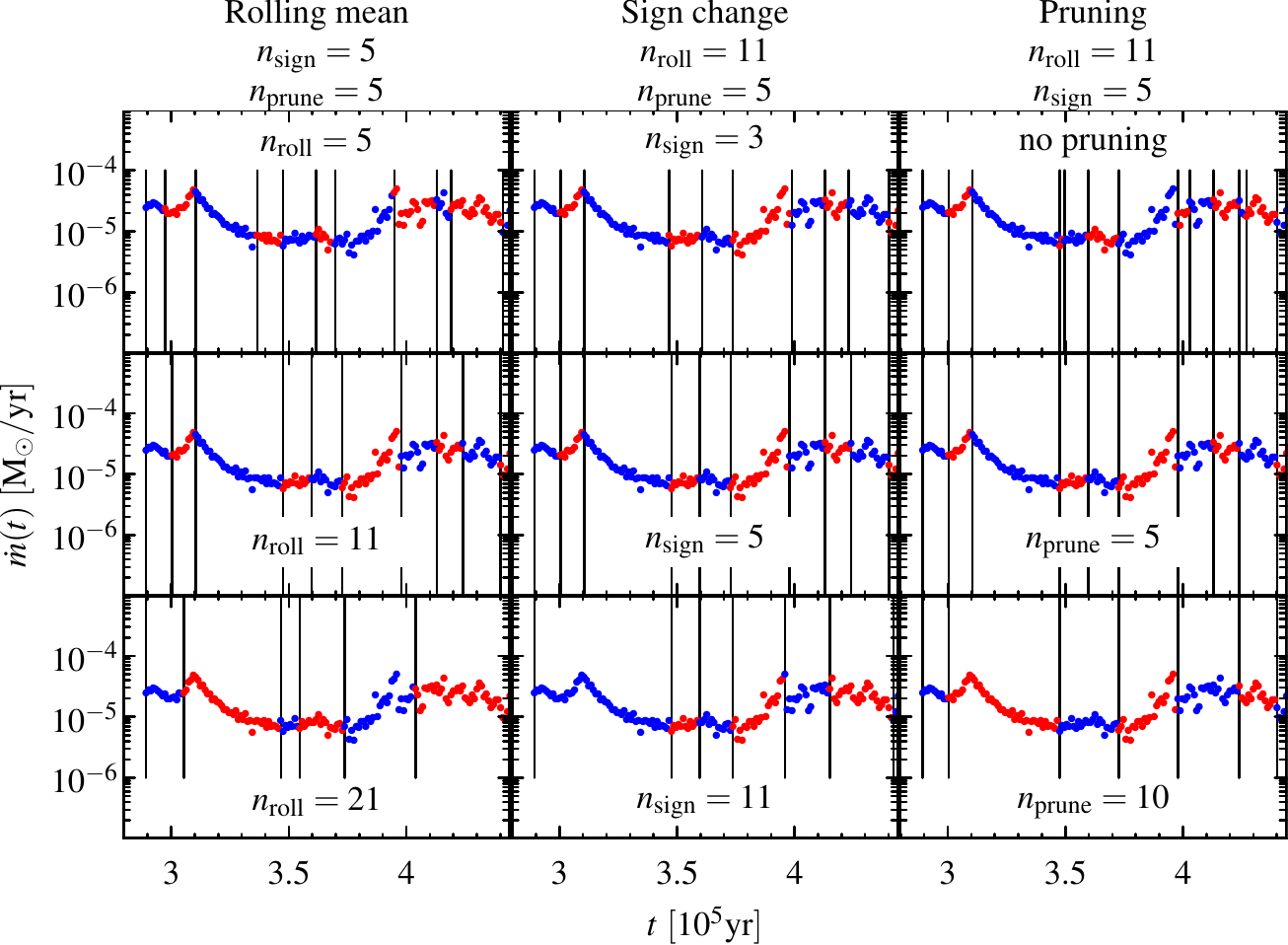}
\end{center}
\caption{\label{figure_fit_example}
Effects of the parameter choice on the episode detection.
The middle row shows the same plot three times for easier comparison.
}
\end{figure}

\subsection{Identification of the episodes}

We determine the episodes from behaviour of the rolling mean accretion rate as a function of time,
\< \bar{\dot{m}} (t_{i+5}) &=& \frac{1}{11} \sum_{k=i}^{i+11} \dot{m} (t_k). \>
The time window used is 11 000 yr, or 11 data points, which we found to be a reasonable compromise between the smoothness of the mean accretion rate and the resolution of the episodes.
The beginning and end of an episode is characterised by a sign change in the numerical derivative of $\bar{\dot{m}} (t_i)$.
We calculate the sign at time $t_i$ from 5 data points with
\< \text{sign} ( t_{i} ) &=& \bar{\dot{m}} (t_{i-2}) - \bar{\dot{m}} (t_{i+2}). \>
If the sign changes from $t_i$ to $t_{i+1}$ then a new episode starts at $t_{i+1}$.
This procedure leads to some very short episodes, which typically last only a few thousand years.
These are only spurious detections.
Therefore we remove them (pruning) by attaching any episode that is shorter than 5000 years (less than 5 data points) to the previous episode.
The colour-coding in Fig. \ref{figure_example_histories} and Fig. \ref{figure_exponents} shows the episodes identified in this way.
Our episode determination finds the main features in the growth history roughly agreeing with what would be found by human eye.

Figure \ref{figure_fit_example} shows the effects of parameter variations in the episode determination algorithm using the first formed strongly accreting sink as an example.
For clarity only a part of the growth history is shown (complete in the right bottom panels of Fig. \ref{figure_example_histories}).
The vertical lines show the boundaries of the episodes for the parameter choice of the respective panel.
There are three parameters in the episode detection algorithm, the number of data points over which the rolling mean runs ($n_\text{roll}$), the number of data points over which the sign change is determined ($n_\text{sign}$), and the maximum length of episodes which are pruned ($n_\text{prune}$).
In each column of Fig. \ref{figure_fit_example} one of the parameters is varied (the used value is given in the panel), whereas for the other parameters our standard choice is used (given on top of the panel).
The panels in the middle row are identical, corresponding to our adopted choice of parameters, and are shown for easier comparison.
For this particular sink the algorithm should find the first three episodes which are a decreasing $\dot{m}$ from initial collapse ($2.9$--$3\times 10^5$ yr) and then rise ($3$--$3.1\times 10^5$ yr) and again decrease ($3.1$--$3.5\times 10^5$ yr) of $\dot{m}$ of the first accretion event.
Up to approximately $3.7\times 10^5$ yr the accretion rate is rather smooth during the episodes, but afterwards there is a larger amount of fluctuations, which makes episode detection more difficult.

Averaging over a shorter period produces more episodes (top left panel), whereas a longer averaging period produces less episodes (bottom left panel).
Changing $n_\text{sign}$ has the same effect.
Without pruning many very short episodes are produced (after $4\times 10^5$ yr, top right panel), but if the pruning length is very long then real episodes are lost (bottom right panel).
Generally, larger values of the parameters give longer episodes but miss some short ones, while smaller parameter values lead to shorter episodes but more spurious detections.
Our choice of parameters is a compromise between the number and length of episodes.

\subsection{Classification of the growth histories}

\begin{figure}
\begin{center}
\includegraphics[width=8.5cm]{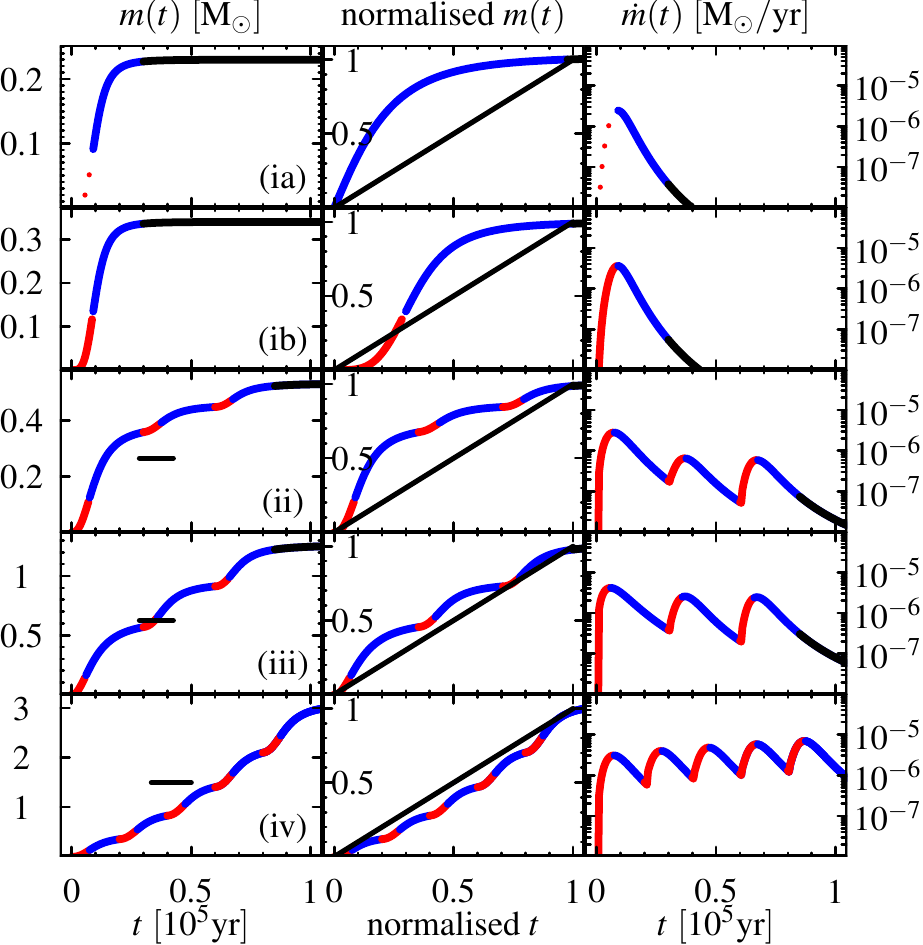}
\end{center}
\caption{\label{figure_episode_schema}
Schema of episodic growth for each class.
The left panels show $m(t)$, which has been normalised in the middle panels.
The right panels show $\dot{m}(t)$.
For class (ia) in the top panel the dotted red curve shows the initial episode, which is not resolved in the simulation.
The black bar in panels (ii), (iii) and (iv) runs from $\frac{1}{3} t_{95\%}$ to  $\frac{1}{2} t_{95\%}$ at half the final sink mass.
}
\end{figure}

\begin{figure}
\begin{center}
\includegraphics[width=7cm]{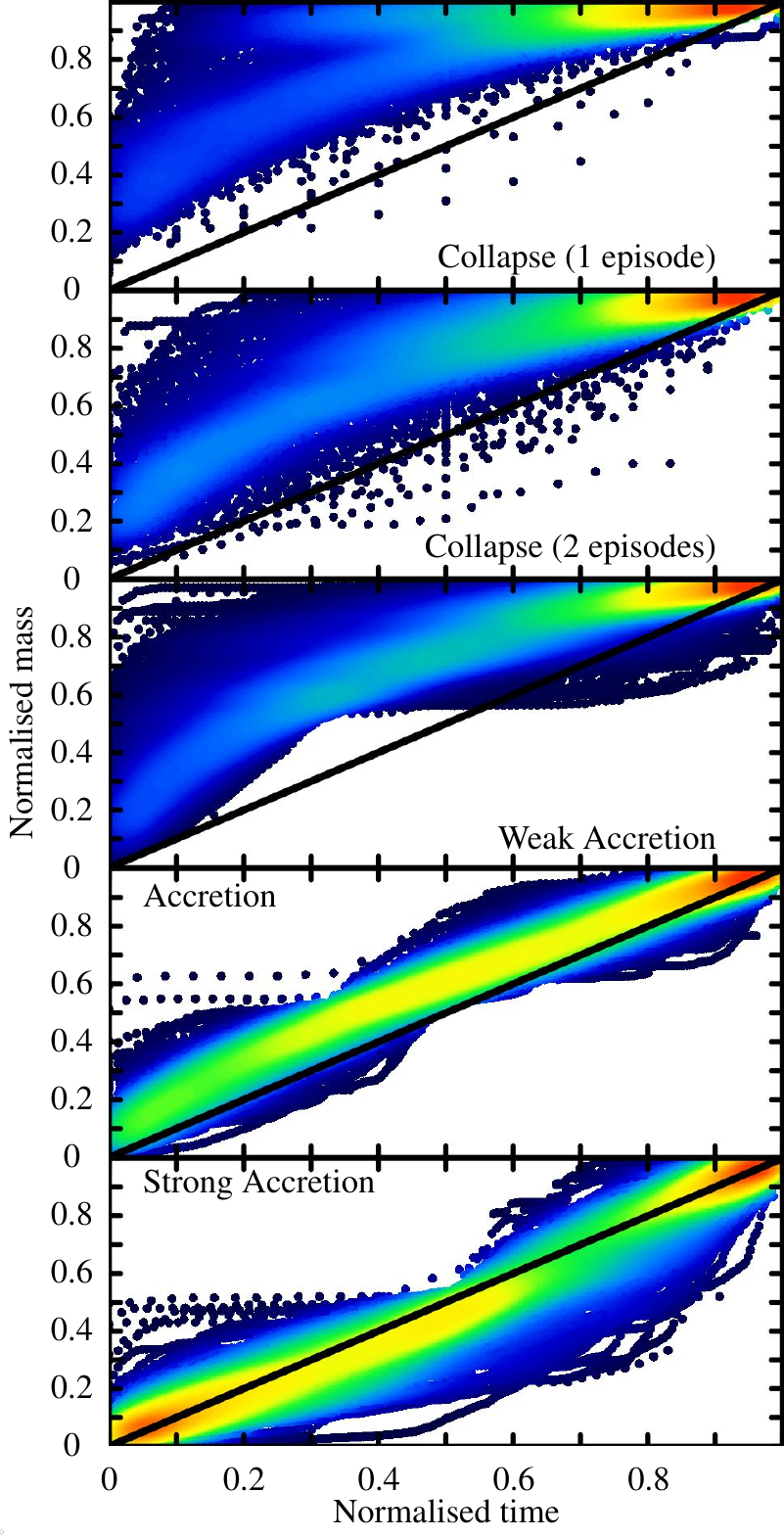}
\end{center}
\caption{\label{figure_characteristic_histories}
Characteristic growth histories of the sinks for each class, which change with increasing amount of accretion  from a concave shape to a convex shape.
Time is normalised as $t/t_{95\%}$ and mass is normalised to the final mass of the sink.
}
\end{figure}

The growth history of a sink  consists of a collapse phase often followed by an accretion phase, consisting of one or many episodes.
The collapse phase falls normally into a single episode, but can sometimes extend over two episodes,
About half of the sinks show significant growth by accretion after the initial collapse phase.
Most sinks show a quiescent phase of very minor mass gain that occurs in the later stages of their evolution, after the initial collapse and any subsequent accretion phases.
Mass growth has effectively stopped when they set in.
Therefore we define the growth time of a sink ($t_{95\%}$) as the time during which 95 per cent of the final mass is assembled.
The parts of the growth histories that fall after $t_{95\%}$ are shown in black in Figure \ref{figure_example_histories}.
Typically the time between $t_{95\%}$ and the end of the simulation covers the final quiescent phase, except for massive sinks, where some accretion is needed for the last 5 per cent of mass gain.
Quiescent phases can also occur between two accretion events before $t_{95\%}$.

This leads us to the following classification scheme for the growth histories of the sinks:\\
\hspace*{1em} (0) {\bf Unresolved collapse:}
Sinks that less than double their initial mass during the simulation.
\\
\hspace*{1em} (ia) {\bf Collapse in 1 episode:}
Sinks that have 75 per cent of their mass gain ($m_\text{end} - m_\text{start}$) in the first episode  and are not class (0).
\\
\hspace*{1em} (ib) {\bf Collapse in 2 episodes:}
Sinks that have 75 per cent of their mass gain within the first two episodes and are not class (0) or (ia).
\\
\hspace*{1em} (ii) {\bf Weak accretion:}
Sinks that achieve at least 50 per cent of their final mass in less than the first 33 per cent of their growth time ($t_{95\%}$) and are not class (0), (ia) or (ib).
\\
\hspace*{1em} (iii) {\bf Accretion:}
Sinks that achieve at least 50 per cent of their final mass in the time between 33 per cent and 50 per cent of their growth time and are not class (0), (ia) or (ib).
\\
\hspace*{1em} (iv) {\bf Strong accretion:}
Sinks that achieve more than 50 per cent of their final mass in the second half of their growth time and are not class (0), (ia) or (ib).

For the classification we first establish whether a sink falls into class (0) or not.
If the sink more than doubles the mass it contains enough data points to proceed with the analysis and classification.
The next step is to establish whether a sink that doubles in mass is collapsing (class (ia) or (ib)) or not.
If the mass gain of a sink is not dominated by the initial collapse then significant amounts of accretion are present and it can be classified as weakly/intermediate/strongly accreting (classes (ii), (iii) or (iv)).
The condition for being in class (ii), (iii) or (iv) are mutually exclusive.
Thus the classification of a sink is unique, it is assigned only one class.
Note that the classification scheme is independent of the final sink mass and only based on the morphology of the growth history.

Sinks in class (0) have more than half of their final mass already at the moment when the sink is formed during the first collapse phase.
Therefore they are collapse-dominated, but their growth history is not resolved by the sink particle, only by the gas particles.
Usually they have very small masses.

\begin{table*}
\caption{\label{table_summary}
Properties of each category. 
Bars denote average values which are quoted with the standard deviation.
Tildes denote the median with the errors corresponding to the quantiles at $\pm 1 \sigma$ (83 and 17 per cent). 
The rows are:
$n$ number of sinks in this class;
`\% all' is the percentage with respect to all sinks; \% with respect only to the mass gaining sinks;
$m$ final mass;
min $m$ the minimum mass of a class, which is affected by outliers. Therefore we give also the 2 per cent quantile ($m_{2\%}$);
max $m$ is the maximum mass of a class;
$m_1/m_0$ mass gain (ratio final/initial mass);
$t_{95\%}$ growth time;
$n_\text{ep}$ number of episodes;
$t_\text{ep}$ duration of episodes;
}
\begin{tabular}{llrrrrrrr}
& 
& 
\hspace{-7em} \rotatebox[origin=rt]{-10}{(0) Unresolved collapse}   &
\hspace{-10em} \rotatebox[origin=rt]{-10}{(ia) Collapse (1 episode) } &
\hspace{-10em} \rotatebox[origin=rt]{-10}{(ib) Collapse (2 episodes)} &
\hspace{-7em} \rotatebox[origin=rt]{-10}{(ii) Weak accretion} &
\hspace{-5em} \rotatebox[origin=rt]{-10}{(iii) Accretion} &
\hspace{-8em} \rotatebox[origin=rt]{-10}{(iv) Strong Accretion}
\\
$n$ & & 
540  & 
441  & 
296  & 
499  & 
338  & 
208  & 
\\
\% all & & 
23  & 
19  & 
13  & 
21  & 
15  & 
 9  & 
\\
\% & & 
 --- & 
25  & 
17  & 
28  & 
19  & 
12  & 
\\
$\bar{m}$ & $\Msun$ & 
0.07 \raisebox{.3ex}{\scriptsize $\pm$ 0.04 } & 
0.23 \raisebox{.3ex}{\scriptsize $\pm$ 0.16 } & 
0.33 \raisebox{.3ex}{\scriptsize $\pm$ 0.25 } & 
0.52 \raisebox{.3ex}{\scriptsize $\pm$ 0.60 } & 
1.25 \raisebox{.3ex}{\scriptsize $\pm$ 1.18 } & 
3.01 \raisebox{.3ex}{\scriptsize $\pm$ 3.63 } & 
\\
$\tilde{m}$ & $\Msun$  & 
0.06 $\genfrac{}{}{0pt}{}{+ 0.04 }{- 0.02 }$  & 
0.19 $\genfrac{}{}{0pt}{}{+ 0.17 }{- 0.10 }$  & 
0.26 $\genfrac{}{}{0pt}{}{+ 0.28 }{- 0.15 }$  & 
0.34 $\genfrac{}{}{0pt}{}{+ 0.50 }{- 0.23 }$  & 
0.92 $\genfrac{}{}{0pt}{}{+ 0.98 }{- 0.48 }$  & 
1.81 $\genfrac{}{}{0pt}{}{+ 3.55 }{- 1.25 }$  & 
\\
$\text{min}\ m$ & $\Msun$  & 
0.01  & 
0.04  & 
0.04  & 
0.03  & 
0.06  & 
0.05  & 
\\
$m_\text{2\%}$ & $\Msun$  & 
0.02  & 
0.05  & 
0.06  & 
0.05  & 
0.09  & 
0.13  & 
\\
$\text{max}\ m$ & $\Msun$  & 
0.36  & 
0.93  & 
1.58  & 
4.70  & 
7.86  & 
30.29  & 
\\
$\bar{m_{1}/m_0}$ & & 
1.49 \raisebox{.3ex}{\scriptsize $\pm$ 0.02 } & 
3.90 \raisebox{.3ex}{\scriptsize $\pm$ 0.05 } & 
5.53 \raisebox{.3ex}{\scriptsize $\pm$ 0.06 } & 
8.51 \raisebox{.3ex}{\scriptsize $\pm$ 0.05 } & 
18.69 \raisebox{.3ex}{\scriptsize $\pm$ 0.09 } & 
52.45 \raisebox{.3ex}{\scriptsize $\pm$ 0.13 } & 
\\
$\tilde{m_1/m_0}$ & & 
1.48 $\genfrac{}{}{0pt}{}{+ 0.31 }{- 0.30 }$  & 
3.38 $\genfrac{}{}{0pt}{}{+ 1.83 }{- 1.10 }$  & 
4.89 $\genfrac{}{}{0pt}{}{+ 2.89 }{- 2.07 }$  & 
5.94 $\genfrac{}{}{0pt}{}{+ 6.59 }{- 3.12 }$  & 
13.10 $\genfrac{}{}{0pt}{}{+ 13.69 }{- 5.71 }$  & 
28.78 $\genfrac{}{}{0pt}{}{+ 49.57 }{- 18.07 }$  & 
\\
$\bar{t_{95\%}}$ & $10^3\ \text{yr}$ & 
26 \raisebox{.3ex}{\scriptsize $\pm$ 34 } & 
34 \raisebox{.3ex}{\scriptsize $\pm$ 31 } & 
50 \raisebox{.3ex}{\scriptsize $\pm$ 30 } & 
115 \raisebox{.3ex}{\scriptsize $\pm$ 59 } & 
121 \raisebox{.3ex}{\scriptsize $\pm$ 65 } & 
160 \raisebox{.3ex}{\scriptsize $\pm$ 75 } & 
\\
$\tilde{t_{95\%}}$ & $10^3\ \text{yr}$ & 
12 $\genfrac{}{}{0pt}{}{+ 36 }{-  9 }$  & 
24 $\genfrac{}{}{0pt}{}{+ 29 }{- 13 }$  & 
43 $\genfrac{}{}{0pt}{}{+ 29 }{- 19 }$  & 
101 $\genfrac{}{}{0pt}{}{+ 73 }{- 42 }$  & 
109 $\genfrac{}{}{0pt}{}{+ 80 }{- 52 }$  & 
152 $\genfrac{}{}{0pt}{}{+ 88 }{- 73 }$  & 
\\
$\bar{n_\text{ep}}$ & & 
 --- & 
2.00 \raisebox{.3ex}{\scriptsize $\pm$ 1.70 } & 
3.16 \raisebox{.3ex}{\scriptsize $\pm$ 1.55 } & 
7.72 \raisebox{.3ex}{\scriptsize $\pm$ 3.71 } & 
7.92 \raisebox{.3ex}{\scriptsize $\pm$ 4.17 } & 
10.60 \raisebox{.3ex}{\scriptsize $\pm$ 5.24 } & 
\\
$\tilde{n_\text{ep}}$ & &
 --- & 
 1 $\genfrac{}{}{0pt}{}{+  2 }{-  0 }$  & 
 3 $\genfrac{}{}{0pt}{}{+  1 }{-  1 }$  & 
 7 $\genfrac{}{}{0pt}{}{+  4 }{-  3 }$  & 
 7 $\genfrac{}{}{0pt}{}{+  5 }{-  3 }$  & 
10 $\genfrac{}{}{0pt}{}{+  6 }{-  6 }$  & 
\\
$\bar{t_\text{ep}}$ & $10^3\ \text{yr}$ & 
 --- & 
23.36 \raisebox{.3ex}{\scriptsize $\pm$ 13.85 } & 
19.19 \raisebox{.3ex}{\scriptsize $\pm$ 10.23 } & 
16.30 \raisebox{.3ex}{\scriptsize $\pm$ 8.75 } & 
16.66 \raisebox{.3ex}{\scriptsize $\pm$ 8.80 } & 
16.21 \raisebox{.3ex}{\scriptsize $\pm$ 8.77 } & 
\\
$\tilde{t_\text{ep}}$ & $10^3\ \text{yr}$ & 
 --- & 
20.11 $\genfrac{}{}{0pt}{}{+ 16.09 }{- 9.05 }$  & 
16.58 $\genfrac{}{}{0pt}{}{+ 11.56 }{- 5.53 }$  & 
14.07 $\genfrac{}{}{0pt}{}{+ 9.06 }{- 5.02 }$  & 
14.08 $\genfrac{}{}{0pt}{}{+ 10.04 }{- 5.03 }$  & 
14.07 $\genfrac{}{}{0pt}{}{+ 9.06 }{- 5.02 }$  & 
\\
$\bar{t_\text{ep}} \bar{n_\text{ep}} $ & $10^3\ \text{yr}$ & 
 --- & 
47  & 
61  & 
126  & 
132  & 
172  & 
\\
$\tilde{t_\text{ep}} \tilde{n_\text{ep}} $ & $10^3\ \text{yr}$ & 
 --- & 
20  & 
50  & 
98  & 
99  & 
141  & 
\\

\end{tabular}
\end{table*}

The typical behaviour of each class is schematically represented in Fig. \ref{figure_episode_schema}, which has the same layout as Fig. \ref{figure_example_histories}.
The left panels show $m(t)$, the middle panels the normalised $m(t)$ and the right panels $\dot{m} (t)$. 
Fig. \ref{figure_characteristic_histories} shows the characteristic growth histories for all sinks in each of the classes (normalised mass vs. normalised time, corresponding to the middle panels of Figs. \ref{figure_example_histories} and \ref{figure_episode_schema}).
This allows us to show all sinks despite their differing masses and growth times in order to see the variations of the growth histories in each class.
The dots are the growth histories of each sink, colour-coded to the point density at their location.

Sinks of class (ia) and (ib) are collapsing cores that do not undergo any significant further accretion.
The signature of a collapse in the $\dot{m}-t$ plot is a very sharp rise of $\dot{m}$ followed by a more gentle decline.
As the increase of $\dot{m}$ can be very fast it is not always completely traced by a sink particle, sometimes the sink is formed only when $\dot{m}$ is already declining.
Then most of the mass is gained in the single episode of declining $\dot{m}$.
This is the case for sinks of class (ia), shown in the top panels of Figs. \ref{figure_example_histories}, \ref{figure_episode_schema} and \ref{figure_characteristic_histories}.
In Fig. \ref{figure_characteristic_histories} the bulk of the sinks behaves as in the schema, but at small values of the normalised time another branch appears in  the upper part.
The top branch is due to (very low mass) sinks that gain a very large fraction of their mass in the collapse, but do not quite reach 95 per cent of their final mass.
A small accretion event is needed to reach the final mass, which can occur a rather long time after the collapse.
An example for this is the first sink formed of class (ia), shown in the top right part of Fig. \ref{figure_example_histories}.

Sinks in class (ib) are formed very early on during the initial collapse and the quickly rising $\dot{m}$ is resolved, so that two episodes are found.
Their behaviour is shown in the second panels from top in Figs. \ref{figure_example_histories}, \ref{figure_episode_schema} and \ref{figure_characteristic_histories}.
Compared to class (ia) the scatter has increased in Fig. \ref{figure_characteristic_histories} and the top branch is not present any more.
Class (ia) can by construction only contain collapsing sinks, whereas class (ib) can contain sinks that had an accretion episode after collapse, if the initial rise of $\dot{m}$ is unresolved.
We did not find a robust and objective way to distinguish between 2-episode collapse and 1-episode plus an accretion episode in class (ib).
Therefore we introduced the split of  collapsing sinks  into classes (ia) and (ib).

Classes (ii), (iii) and (iv) contain sinks that underwent increasing magnitudes of accretion and have more episodes than the sinks in classes (ia) and (ib).
Accretion does not proceed in a smooth way, perhaps with some scatter in the accretion rates, but as a sequence of accretion events after the initial collapse.
This is particularly visible in the bottom panels for $\dot{m}(t)$ of Fig. \ref{figure_example_histories} which has a zigzag shape from the sharp rise and decline of $\dot{m}$ during the secondary `accretion collapses'.
However, our classification scheme is for those sinks not based on the $\dot{m}(t)$, but on the time when the majority of mass is acquired.
The lower three panels of Fig. \ref{figure_episode_schema} for accreting sinks  show a black bar which runs from $\frac{1}{3} t_{95\%}$ to $\frac{1}{2} t_{95\%}$ at half of the final mass.
$m(t)$ for the weakly accreting sinks (class (ii)) runs to the left of the bar, for accreting sinks (class (iii)) it goes through the bar, and strongly accreting sinks (class (iv)) have $m(t)$ that goes below the bar.
The change of $m(t)$ from convex (class (ii)) via linear (class (iii)) to concave (class (iv)) is easier to identify than the change in behaviour of $\dot{m} (t)$ (see Fig. \ref{figure_example_histories}).
In the lower three panels of Fig. \ref{figure_characteristic_histories} this change of morphology is well visible.

\subsection{Properties of the sink classes}

Table \ref{table_summary} gives some characteristic quantities for each sink class where bars denote the mean and tildes  the median.
As the distributions of these quantities are very skewed the standard deviation can be larger than the average.
Therefore we also give with the median mass the quantiles corresponding to $\pm 1 \sigma$ (83 and 17 per cent).
About 23 per cent of all sinks do not double their mass during their stay in the simulation.
Of those that significantly gain mass 40 per cent only collapse (classes ia and ib), $\approx$ 30 per cent show weak accretion (ii), $\approx$ 20 per cent are accretion-dominated (iii) and $\approx$ 10 per cent have strong accretion (iv).
There is a steady increase in the mean and median mass of each classes (but note that mass is not a criterion for classification).
The minimum mass in each class is probably affected by some outliers due to misclassification, hence we also provide the 2 per cent quantile.
Collapsing sinks have on average a mass of $\approx 0.3\ \Msun$, but strongly accreting sinks are a factor of 10 more massive.
Similarly, the average mass gain (ratio of final and initial mass, $m_1/m_0$) ranges from a factor of $\approx 4$ up to a factor of $\approx$ 50.
With increasing amount of accretion also the growth time increases, as well as the number of episodes ($n_\text{ep}$).
The duration of the episodes ($t_\text{ep}$), however, is fairly constant for each class and not much affected by the presence and amount of accretion.

\section{Accretion rates}\label{sec_accretion_rates}

\subsection{Mass dependence of the accretion rates}

\begin{figure}
\begin{center}
\includegraphics[width=7cm]{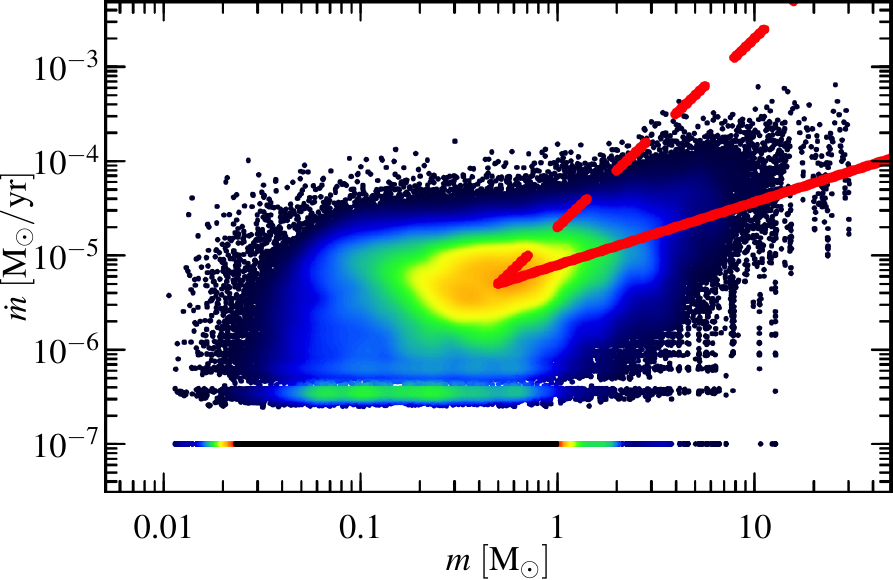}
\end{center}
\caption{\label{figure_accretion_rates}
Plot of the accretion rates vs. m for all sinks at all times of the simulation, colour-coded to the point density.
Lines are for $\dot{m} \propto m^2$ (dashed) and $\dot{m} \propto m^{2/3}$ (solid).
Time intervals with no accretion have been assigned a fiducial accretion rate of $10^{-7}\ \Msun/\text{yr}$. 
}
\end{figure}

\begin{figure*}
\includegraphics[width=14cm]{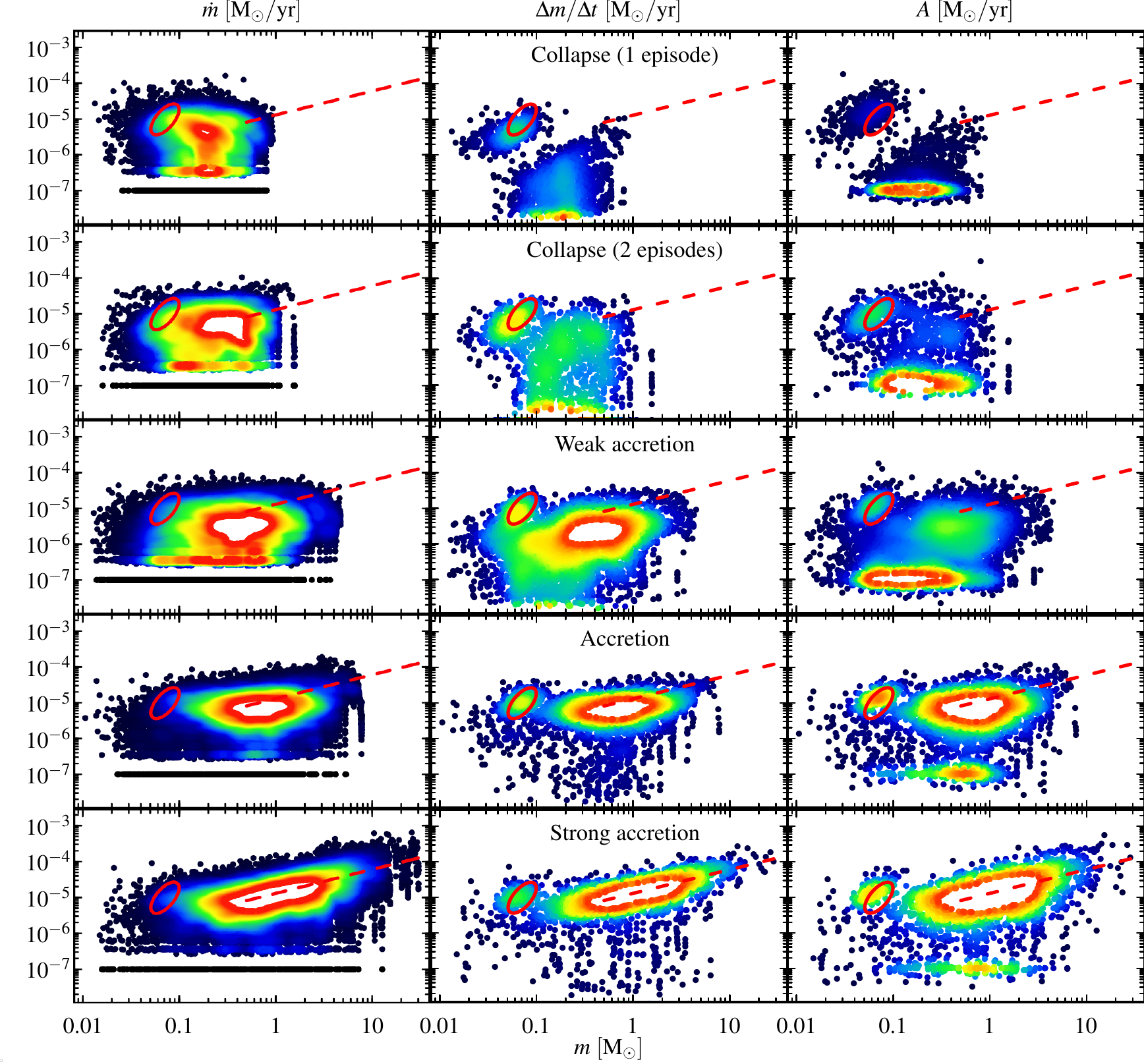}
\caption{\label{figure_accretion_rates_classes}
Left panel: Accretion rates as a function of sink mass for each class. 
Middle panel: Episodic accretion rate, defined as mass gain during an episode, $\Delta m$, per duration of episode, $\Delta t$, as function of the sink mass at the beginning of the episode. 
Right panel: Scaling constant $A$ of a fit $\dot{m}=A \text{e}^{bt}$ for each episode versus the sink mass at the beginning of the episode. 
Three groups of episodes become visible, a collapse type (red ellipse), an accreting type where $\dot{m} \propto m^{2/3}$ (dashed line), and quiescent episodes below $\approx 10^{-6}\ \Msun \text{yr}$.
}
\end{figure*}

Fig. \ref{figure_accretion_rates} shows the accretion rate as a function of mass for all sinks in the simulation at all sampling times.
The dots are colour-coded to the point density at their location.
Accretion events can be discrete because of the discrete modelling of the gas density so that there appear stripes of points at the bottom of the plot.
This corresponds to the accretion of a one single, two etc. gas particles to the sink during the sampling time interval. 
Sampling intervals without any accretion of a gas particle, which are not uncommon, are shown as the stripe at $10^{-7}\ \Msun/\text{yr}$.

Below $\approx 0.5\ \Msun$ the accretion rates appear to be independent of the sink mass, although there is a slight trend of a decrease of $\dot{m}$ with $m$.
Above $\approx 0.5\ \Msun$ the accretion rates increase with mass, following approximately $\dot{m} \propto m^{2/3}$.
This is predicted for accretion in gas-dominated potentials \citet{BonnellBateClarke-2001a,BonnellClarkeBate-2001a}.
Certainly the sinks do not follow classical Bondi--Hoyle accretion $\propto m^2$, which corresponds to the dashed line.
Besides the mass scaling there is a considerable scatter in the accretion rates, spanning more than an order of magnitude.
Furthermore, there are the `icicles' in the $\dot{m}$--$m$ plot, strands of decreasing accretion rates at the same mass, which belong to the same sink.
These are particularly visible at large masses.

The mass dependence of the accretion rates has been studied by several authors.
For small masses ($m < 0.5 \ \Msun$)
\citet{BateBonnell-2005a,Bate-2009a,Bate-2009b,Bate-2012a}
find no mass dependence of the {\it time-averaged} $\bar{\dot{m}}$, which is consistent with Fig. \ref{figure_accretion_rates}.
\citet{OffnerKleinMcKee-2009a} fit also the {\it  time-averaged} accretion rates and find $\bar{\dot{m} } \propto m^{0.64}$ without radiative transfer and $\bar{\dot{m}} \propto m^{0.92}$ including radiative transfer in the calculation.
\citet{DibShadmehriPadoan-2010a} reports that  in the simulations of \citet{SchmejaKlessen-2004a} the {\it final} masses scale with the {\it peak} accretion rate as $\dot{m}_\text{peak} \propto m_\text{final}^{0.65}$.
However, as very likely in these simulations sink growth is episodic as in ours, the time-averaged or peak accretion rate may not necessarily give the appropriate mass scaling.

\subsection{Accretion rates of the individual classes}

With the grouping of the sinks in various growth classes we are able to disentangle Fig. \ref{figure_accretion_rates}.
This is done in Fig. \ref{figure_accretion_rates_classes}, which shows $\dot{m}$ vs. $m$ for each class individually.
The left column gives $\dot{m}$-$m$ sampled at 1000 yr intervals where the `icicles' of exponentially decaying accretion rates are well visible.
Again we add $10^{-7}\ \Msun/\text{yr}$ to the accretion rate in order to be able to show episodes of extremely low or no accretion in the logarithmic plot.
As the episodic accretion produces a large spread in $\dot{m}$ we show in the middle column of Fig. \ref{figure_accretion_rates_classes} the average accretion rates during each episode.
This is given by the fraction of mass accreted during an episode, $\Delta m$, divided by the duration of the episode, $\Delta t$.
The mass coordinate is the mass at the beginning of the episode.
Here many of the `icicles' have vanished and the scatter is reduced.
The average accretion rates $\Delta m/\Delta t$ during an episode depend on the length of the episode. 
In order to assess the length dependence we show in the right panel of Fig. \ref{figure_accretion_rates_classes} the scaling constant $A$ of a fit $\dot{m} = A \e^{b t}$ as a function of $m$ at the beginning of an episode.
$A$ shows the same behaviour as $\dot{m}-m$ or $\Delta{m}/\Delta{t} - m$, in particular the same $m^{2/3}$ scaling.
Compared to  $\Delta{m}/\Delta{t} - m$ there is more scatter in the distribution of $A$.

In the middle and left column of Fig. \ref{figure_accretion_rates_classes} it is very evident that there are three types of episodes: initial collapse, accretion and quiescent.
The initial collapse is located at $\approx 0.08\ \Msun$ and $10^{-5}\ \Msun/\text{yr}$, marked by an ellipse.
This phase is well separated from the two others. 
For the strongly collapsing sinks in one phase the initial collapse is followed mainly by quiescent phases, although some subsequent accretion occurs on a small level.
The quiescent phases are located below $10^{-6}\ \Msun/\text{yr}$ at the bottom of the panels.
Accretion is at a higher $\dot{m}$, steadily increasing from panel to panel downwards.
Finally, for the accreting or strongly accreting sinks (class iii/iv) a very clear mass dependence of the accretion rates becomes evident, scaling $\propto m^{2/3}$ (dashed line).
This scaling is expected from accretion in a gas-dominated potential \citep{BonnellBateClarke-2001a,BonnellClarkeBate-2001a}.
As accretion increases quiescent phases become more and more rare.

The instantaneous accretion rates in the left column of Fig. \ref{figure_accretion_rates_classes} show similar trends as the in middle and  right column.
However, due to the exponential time-dependence of the accretion rates the features are smeared out, particularly the collapse phase.
Also, this leads to a very strong population of very small accretion rates.

\subsection{Fluctuations in the accretion rates}\label{subsection_accretion_individual}

\begin{figure}
\begin{center}
\includegraphics[width=6cm]{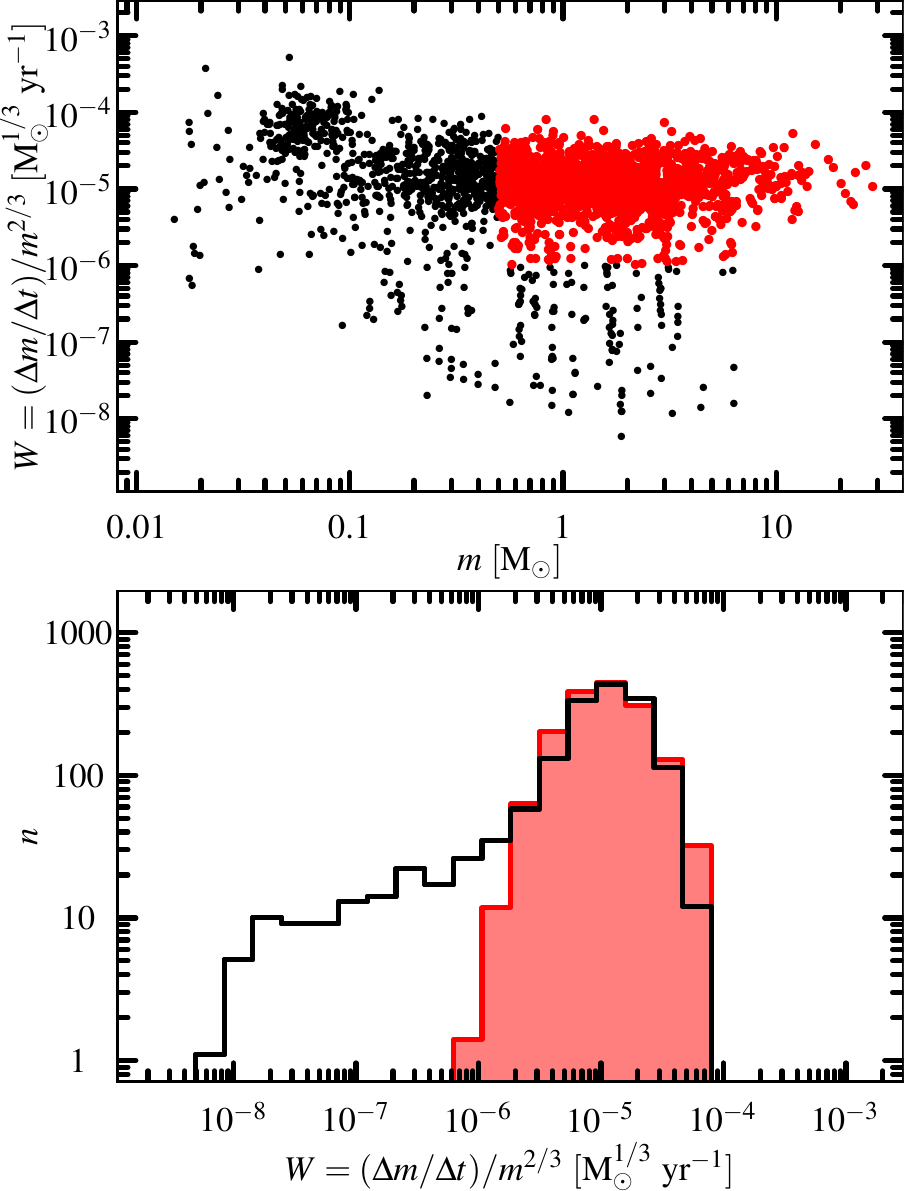}
\end{center}
\caption{\label{figure_mdot_distribution}
Fluctuations $W$ in the accretion rates.
Scatter plot as a function of mass in the upper panel.
The lower panel shows a histogram of $W$ for $m > 0.5\ \Msun$ with a lognormal fit for $W$ (shaded histogram) during accreting episodes ($W > 10^{-6}\ \Msun^{1/3} \text{yr}^{-1}$, i.e. the red dots in the upper panel).
Below $\approx 10^{-6}\ \Msun^{1/3} \text{yr}^{-1}$ the distribution of $W$ is dominated by quiescent episodes.
}
\end{figure}

The time-averaged accretion rates show besides their mass dependence a large scatter, which can be written as
\< \frac{\Delta m}{\Delta t} &=& m^{2/3} W,\>
where $W$ is a fluctuating quantity generating the scatter in the accretion rates.
In order to estimate the amount of scatter we need to determine the distribution of $W$ with its parameters.
The strongly accreting sinks have the widest mass range over which they follow $\dot{m} \propto m^{2/3}$.
Therefore we analyse the distribution of $W$ for the strongly accreting sinks, shown in Fig. \ref{figure_mdot_distribution}.
The top panel shows $W$ as a function of the sink mass.
Like in Fig. \ref{figure_accretion_rates_classes} the three types of episodes are visible, collapsing, quiescent and accreting.
As the mass dependence for $W$ is removed, the distribution of $W$ is constant in mass.
It appears that during accreting episodes (marked in red) $W$ follows a lognormal distribution,
\< p (W) &=& \frac{1}{W} \frac{1}{\sqrt{2 \upi} \sigma_{W}}  \e^{\displaystyle - \frac{1}{2} 
\frac{ \( \log W - \mu_{W} \)^2}{\sigma_{W}^2} }. \>
Accreting episodes are selected by requiring that $W > 10^{-6}\ \Msun^{1/3} \text{yr}^{-1}$ and $m > 0.5\ \Msun$ (shown as red dots).
A maximum likelihood fit of the parameters of a lognormal distribution gives  $\mu_{W}= -11.44$ and  $\sigma_{W}= 0.74$.

In order to test the goodness of a lognormal fit we show in the bottom panel of Fig. \ref{figure_mdot_distribution} a histogram of $W$, selecting all those with $m > 0.5\ \Msun$ (including quiescent episodes).
The red shaded histogram shows a lognormal distribution with the estimated parameters.
Above $10^{-6}\ \Msun^{1/3} \text{yr}^{-1}$ the distribution of $W$ agrees reasonably with the lognormal fit, but below $10^{-6}\ \Msun^{1/3} \text{yr}^{-1}$ there is an excess of small $W$.
The low-$W$ part of the distribution contains mainly quiescent episodes, and is thus not representative for the distribution of $W$ during accreting episodes.
Therefore the assumption of a lognormal distribution of $W$ during accreting episodes is justified by the data.

\section{Location of the classes in the sink mass function and distribution of initial collapse masses}\label{sec_mass_fn}

\begin{figure}
\begin{center}
\includegraphics[width=7cm]{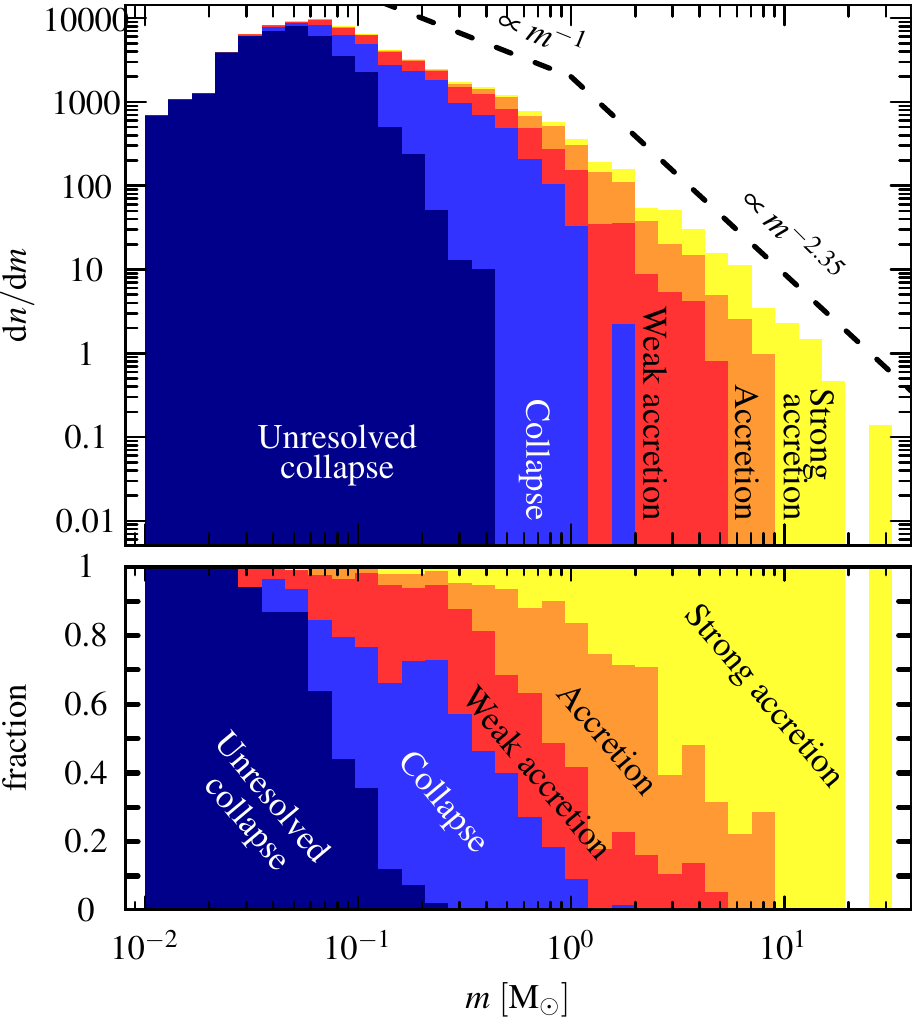}
\end{center}
\caption{\label{figure_class_imf}
Top panel:
Sink mass function at the end of the calculation where the contribution of each category is shown. 
Bottom panel: ratio to which each class contributes at a given mass range.
}
\end{figure}

\begin{table*}
\caption{
Results from the mass function fits.
$\alpha$, $\beta$ and $\mu$ are the parameters of the $L_3$ IMF.
$\alpha$ is the high-mass exponent and $\gamma$ the low-mass exponent.
$m_\gamma$ and $m_\alpha$ are the masses below or above which the $L_3$ IMF follows power-laws.
The mode refers to the maximum of $L_3$ in linear units (as Fig. \ref{figure_mass_histograms}), whereas $m_\text{Peak}$ refers to the maximum in logarithmic units.
}
\begin{tabular}{lrrrrrrrrr}
Class & $\alpha$ & $\beta$ & $\mu$ & $\gamma$ & $m_\gamma$ & $m_\alpha$ & Mode & $m_\text{Peak}$ & Average mass \\
All sinks, collapse mass  & 
3.36  & 
1.67  & 
0.18  & 
-0.58  & 
0.078  & 
0.42  & 
0.09  & 
0.15  & 
0.19 \\
Collapsing sinks, collapse mass  & 
2.80  & 
2.91  & 
0.07  & 
-2.44  & 
0.021  & 
0.20  & 
0.06  & 
0.09  & 
0.15 \\
Accreting sinks, collapse mass  &
4.19  & 
1.48  & 
0.28  & 
-0.52  & 
0.150  & 
0.53  & 
0.15  & 
0.22  & 
0.23 \\
All sinks, final mass  & 
1.93  & 
10.57  & 
0.01  & 
-7.90  & 
0.002  & 
0.11  & 
0.06  & 
0.15  & 
0.67 \\
Collapsing sinks, final mass  & 
2.53  & 
3.53  & 
0.05  & 
-2.88  & 
0.015  & 
0.20  & 
0.06  & 
0.10  & 
0.19 \\
Accreting sinks, final mass  & 
2.47  & 
1.93  & 
0.67  & 
-0.37  & 
0.173  & 
2.60  & 
0.18  & 
0.64  & 
1.25 \\
\end{tabular}

\end{table*}

\begin{figure*}
\begin{center}
\includegraphics[width=16cm]{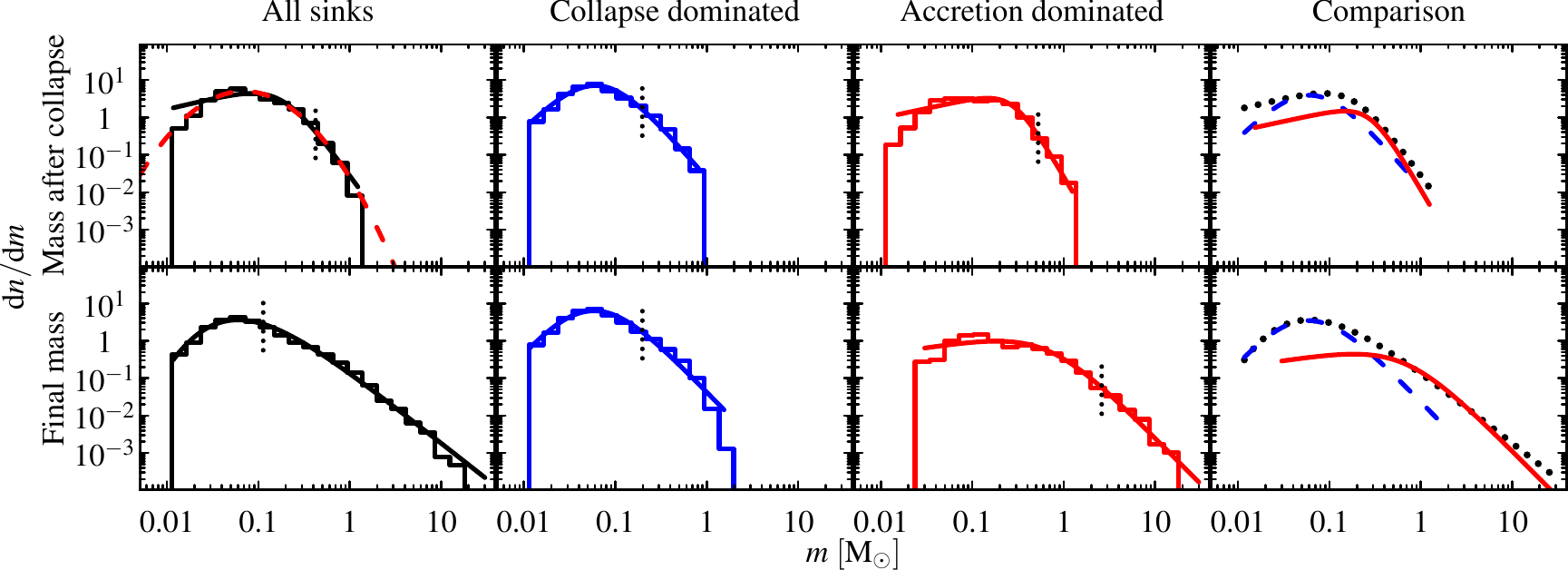}
\end{center}
\caption{\label{figure_mass_histograms}
Distribution of the sink masses after initial collapse (upper row) and at the end of the simulation (lower row).
The curves are $L_3$ fits to the data, where the dotted lines indicate the start of the power-law tail.
In the leftmost upper panel the red dashed line gives a lognormal fit.
In the right column the dashed curve for the collapse dominated sinks and the solid curve for the accretion dominated sinks have been scaled by the relative number of these classes for comparison with the mass functions of all sinks (dotted curve).
}
\end{figure*}

Our classification of the growth histories is independent of the final mass and the growth time of each sink.
It merely depends on the behaviour and shape of the growth history.
Within the framework of gas-dominated competitive accretion  \citep{BonnellBateClarke-2001a,BonnellClarkeBate-2001a}.
 it is expected that massive stars gain most of their mass via accretion of initially unbound gas after the collapse of the initial core.
As our classification groups the sinks according to the amount of post-collapse accretion that they undergo we would expect that the strongly collapsing sinks without much accretion populate the low-mass range, whereas the strongly accreting sinks have large masses.
This is exactly what we find.
Fig. \ref{figure_class_imf} shows in the top panel a histogram of the final sink masses divided into the classes.
The amount of accretion in a sink's growth history increases with increasing final sink mass.
The overall shape of the final sink mass function has three main components, a lognormal-like low mass part which passes at $\approx 0.1\ \Msun$ into a power law  $\propto m^{\approx -1}$ followed by another power law $\propto m^{\approx -2.5}$ above $\approx 1\ \Msun$.

The lower panel of Fig. \ref{figure_class_imf} shows the contribution of each class in each mass bin.
Below $\approx 0.1\ \Msun$ sinks do not significantly grow after their formation.
Between $\approx 0.1\ \Msun$ and $\approx 0.5\  \Msun$ the mass gain is dominated by collapse.
Above $\approx 0.5\ \Msun$ most sinks have undergone a considerable amount of accretion.

Fig. \ref{figure_mass_histograms} shows the effects of initial collapse and accretion on the mass function.
The top panels of Fig. \ref{figure_mass_histograms} show the mass functions after the initial collapse (i.e. after the first episode), whereas the bottom panels show the mass functions at the end of the simulation.
As we are interested in the origin of the high-mass power-law part of the initial mass function we use the $L_3$ IMF from \citet{Maschberger-2013a} as fit function for comparison with the histogram,
\< p_\text{L3} (m) &\propto& \( \frac{m}{\mu} \)^{-\alpha} \( 1 + \( \frac{m}{\mu}\)^{1-\alpha} \)^{-\beta}.\>
This functional form has a lognormal-like main body with power-law tails at both the low-mass and the high-mass end ($\propto m^{-\alpha}$ in the latter case).
The dotted vertical lines in Fig. \ref{figure_mass_histograms} are positioned at $m_{\alpha}$, the mass at which mass function is effectively a power-law in the high-mass range.

The first column comprises all classes of sinks, including class (0) having no significant mass gain.
Their mass after initial collapse is their final mass, which we use in the lower panel.
After the initial collapse the mass function is roughly lognormal (dashed line), perhaps following a steep power law at the high-mass end.
The final mass function in the lower left panel shows a distinct power law at the high mass end.

The group of the collapse-dominated sinks (i.e. classes (0), (ia) and (ib)), shown in the second column, does not show a large variation between the initial collapse and the final mass function.
There appears to be a high-mass power-law tail, presumably truncated around a few \Msun.
In contrast, the group of accreting sinks (classes (ii), (iii) and (iv)) in the third column shows a strong difference between initial collapse and final mass function.
The final mass function is shifted to higher masses, widened, and shows a power-law tail at high masses.

The last column of Fig. \ref{figure_mass_histograms} shows the $L_3$ fit for the groups, dotted for all sinks, dashed for the collapse group and solid for the accretion group.
For the  collapse and the accretion group, the mass functions have been scaled by the relative number in the group.
The initial collapse mass functions are also very similar, although the accreting group seems to be at slightly higher masses with a steeper power law.
An interesting effect appears in the final mass function of all sinks. 
This mass function has two power-law regimes above the lognormal part.
The power law $\propto m^{\approx - 1}$ in the mass range $0.1-1\ \Msun$ originates from the superposition of the `massive' collapsing sinks and the `low-mass' accreting sinks.
Above $\approx 1\ \Msun$ the accreting sinks dominate and provide the power-law $\propto m^{\approx - 2.35}$.
In the $L_3$ fit the whole mass range from 0.1 to 30 \Msun\ is fitted by a single power law, which leads to the smaller high-mass exponent $\alpha = 1.93$ and the extreme values for $\beta$ and $\mu$.

The collapse mass should correspond to the mass that a core had before collapsing to a sink.
Thus we can compare the collapse masses to theoretical predictions for core masses.
The theories of \citet{PadoanNordlundJones-1997a},  \citet{PadoanNordlund-2002a}, and \citet{HennebelleChabrier-2008a,HennebelleChabrier-2009a,HennebelleChabrier-2013a} predict all mass functions following a lognormal distribution at small masses and a power-law at large masses. 
For each of the theories the power-law exponent is different, and, as the distributions of $m_\text{coll}$ do not extend very far in this regime we do not attempt a comparison with any of the theories.
However, in the lognormal part the theories are in agreement, all predicting that the mass function should have logarithmic width $\sigma_m=\sigma_\rho/2$, half the width of the gas density distribution.
This is a simple consequence of the (inverse) square root dependence of Jeans mass on density, so that a log-normal density field maps onto a log-normal mass function with half the width.

We find that the distribution of $m_\text{coll}$ of all sinks is reasonably fitted by a lognormal distribution (dashed line in Figure \ref{figure_mass_histograms}, middle panel first column) with $\sigma_m=0.82 $.
If we assume that $\dot{m} \propto \rho$ then we might expect that the distribution of accretion traces the distribution of density and hence, as a first guess that $ \sigma_\rho = \sigma_W$. 
Proceeding under this assumption, we have $\sigma_W =0.74$ and might therefore expect that $\sigma_m$ is half that value.
Its measured value ($0.82$) is thus far too large to be consistent with this chain of assumptions. 
This discrepancy might imply that the core masses cannot be deduced from the form of the density field; on the other hand it may simply imply that $\sigma_W$ is not after all a good measure of $\Delta_W$ (and indeed it is to be expected that the distribution of densities probed by accreting sinks is likely to be narrower than the entire range of densities in the simulation).
We do not investigate this issue further here.

\section{Distribution of growth times}\label{sec_growth_times}

\begin{figure}
\begin{center}
\includegraphics[width=7cm]{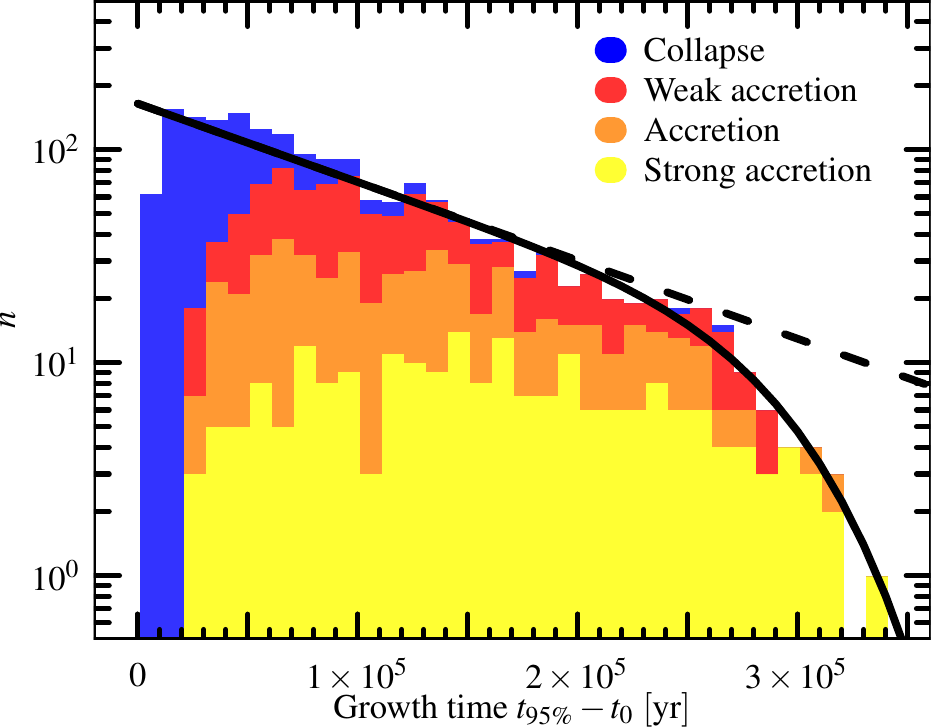}
\end{center}
\caption{\label{figure_histogram_growth_times}
Histogram of the distribution of the growth times colour coded according to growth category.
The curve is an exponential distribution, $p \propto \exp(-t/\theta)$ without (dashed) and with a tapering at large times.
Note that the strongly accreting sinks have an effectively constant distribution of growth times.
}
\end{figure}

\begin{figure}
\begin{center}
\includegraphics[width=7cm]{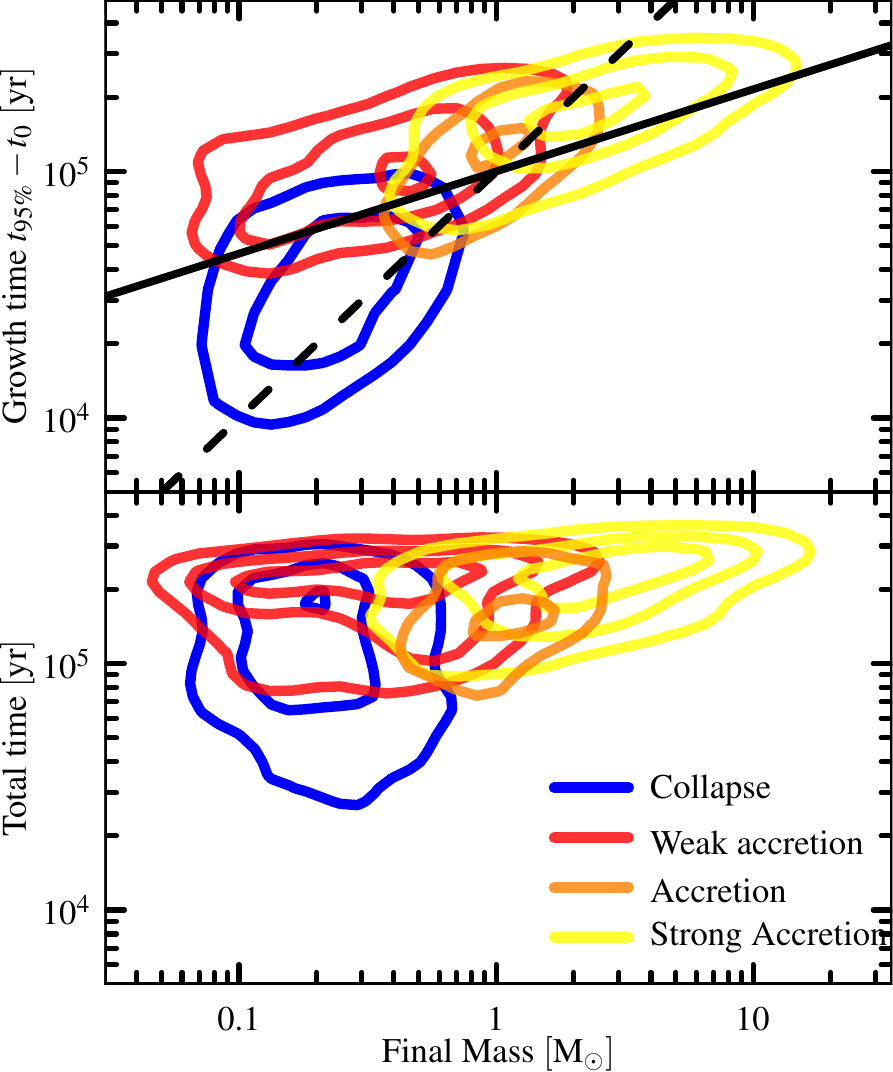}
\end{center}
\caption{\label{figure_growth_times}
Contour plot of the growth times ($t_{95\%}$) vs final mass of the sinks, split by growth category in the upper panel and in the lower panel of the total times the sinks spent in the simulation.
The solid line shows $t \propto m^{1/3}$ and the dashed line a linear relation between the growth time and the final mass.
}
\end{figure}

Fig. \ref{figure_histogram_growth_times} shows the distribution of growth times (the time from sink formation until 95 per cent of mass is reached), divided into categories.
Including all sinks, the growth time follows over a wide range an exponential distribution,
\< p(t) &\propto & \e^{ -\frac{t}{\theta}}, \>
(dashed line) but very large times are underpopulated as the calculation covers only a finite duration.
The decay can be modelled with a `tapered' exponential distribution,
\< p(t) &\propto & \e^{ - \frac{t}{\theta} - \( \frac{t}{\tau} \)^\alpha }, \>
where
$\theta= 1.2 \times 10^5\ \text{yr}$,
$\tau = 3 \times 10^5\ \text{yr}$,
and  $\alpha = 7.1$.
This is shown as solid curve.
The fact that the large-$t$ part of the exponential distribution is missing may have an effect on theories that build on stars growing for a long time to populate the high-mass tail of the mass function \citep[e.g.][]{BasuJones-2004a,BateBonnell-2005a,Myers-2009a}.
It could either steepen the mass function or lead to a truncation at the massive end \citep[which, interestingly, is observed in this simulation, see][]{MaschbergerClarkeBonnell-2010a}.

Not all sink categories follow the same distribution of growth times.
Collapsing sinks have growth times shorter than accreting sinks.
The strongly accreting sinks have a growth times which are only very weakly exponentially distributed, following more a constant distribution.

The amount of accretion correlates with the final mass of the sink, so that there is also a correlation between the growth time and the final masses of a sink.
This is visible in the top panel of Fig. \ref{figure_growth_times}, which shows contours for the different growth classes.
Strongly collapsing sinks reach 95 per cent of their final mass within a few $10^4$ yr,  (weakly) accreting sinks within around $10^5$ yr and strongly accreting sinks in 1--3 $\times 10^5$ yr.

\citet{BateBonnell-2005a}, and  \citet{Bate-2009a,Bate-2012a} find a linear relation between the growth time and the final mass of the sink. (dashed line in the top panel of Fig. \ref{figure_growth_times}).
This is a consequence of $\dot{m} = \text{const}$, supporting the model by \citet{BateBonnell-2005a}.
For  $\dot{m} \propto m^{2/3}$ we expect that $ t \propto m^{1/3}$, which is shown as solid line in  the top panel of Fig. \ref{figure_growth_times}.
This relation seems to follow more closely the contours of the accreting sinks than a linear relation between $m$ and $t$.

The relation between the amount of accretion and the growth time of the sink is not only due to the fact that the low-mass sinks are formed last in the simulation and simply do not have enough time to accrete.
The lower panel of Fig. \ref{figure_growth_times} shows contours for the total lifetime of the sinks, from formation until the end of the simulation.
Although some of the strongly collapsing sinks have a total lifetime less than $10^5$ yr, there is a large fraction of them which spend enough time in the simulation to have undergone at least some accretion.

\section{What determines the shape of the IMF?}\label{sec_imf}

\begin{figure}
\begin{center}
\includegraphics[width=8.5cm]{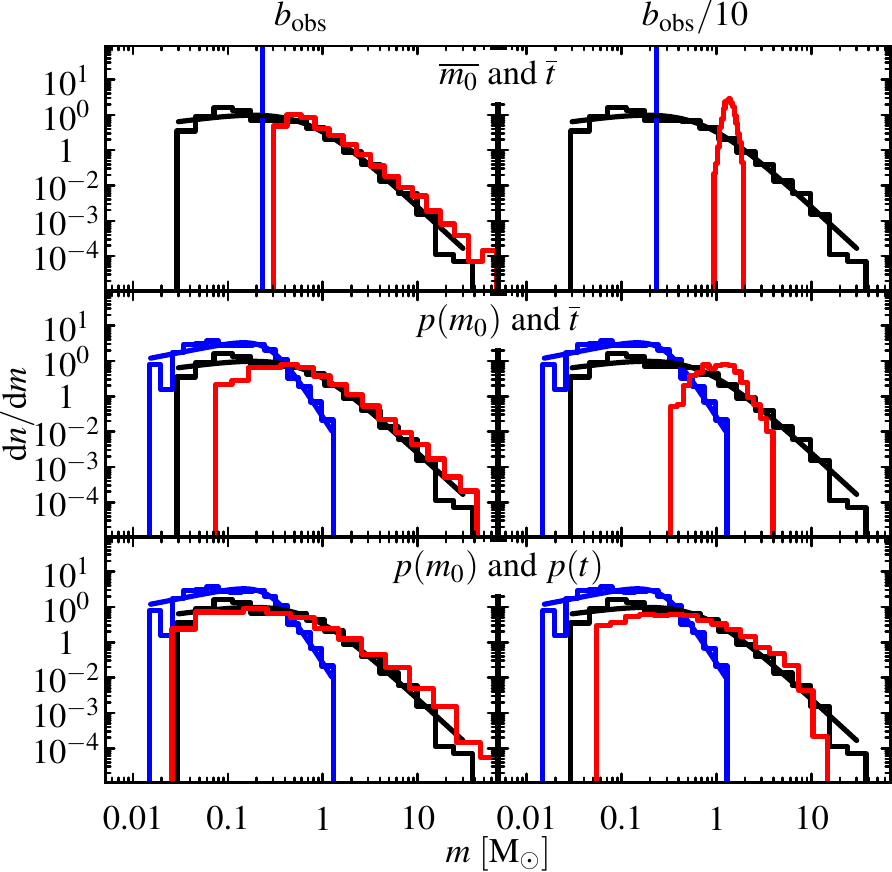}
\end{center}
\caption{\label{figure_imf_fit}
Mass distribution functions obtained by semi-analytical modelling (red histograms) in comparison to the distribution of the final sink masses of the strongly accreting sinks (black histograms).
The blue histogram shows the distribution of collapse masses.
On the left side the amount of fluctuations is of the same magnitude as in the numerical calculation, on the right side reduced by a factor of ten.
The top panels have no distribution of the initial masses and growth times,
the middle panels only a distribution of initial masses and the bottom panels a distribution of initial masses and growth times.
}
\end{figure}

There are three factors that influence the mass function of the accreting sinks: 1) the fluctuations of the accretion rates, 2) the distribution of initial collapse masses and 3) the distribution of the growth times.
But is there one of the factors that is primarily responsible for the shape?
In the simulation all factors occur together, none can be singled out.
Therefore we resort to semi-analytically modelling the growth process in the simulation.
In this way each of the factors can be controlled independently and switched off or on to see which effect it has.

Accretion $\propto m^{2/3}$ with fluctuations in the accretion rates can be described by a stochastic differential equation, which has to account for the lognormal distribution of the accretion rates.
Such stochastic differential equations have been investigated in \citet{Maschberger-2013b}.
In our case the stochastic differential equation is 
\< \d m = m^{2/3} \d \text{iG}_{a,b}, \>
where $\d \text{iG}_{a,b}$ is a random variate from the inverse Gaussian distribution with mean $a$ and standard deviation $b$, describing the fluctuations.
As a lognormally distributed fluctuation term does not allow for an analytical solution we approximate the lognormal distribution by an inverse Gaussian distribution,
\< p_\text{invGauss} (x; \nu,\lambda) &=& \( \frac{\lambda}{2 \upi x^3} \)^{\frac{1}{2}} \e^{- \frac{\lambda (x-\nu)^2 }{2 \nu^2 x } }, \>
which has expectation value  $E(x) = \nu$ and  variance  $\text{Var} (x) = \nu^3/\lambda$.
The infinitesimal fluctuations are then
\< \d \text{iG}_{a,b} & \text{distributed as} & p_\text{invGauss} \( \nu = a \d t , \lambda = \frac{a^3}{b^2} (\d t)^2 \). \>
After growing for a time $t$, starting with a seed mass $m_0$, a sink particle has the mass
\< m(t) &=& \( (1-\alpha)  \( \frac{m_0^{1-\alpha}}{1-\alpha} +  \text{iG}_{a,b,t} \)  \)^{\frac{1}{1-\alpha}},
\>
where
\< \text{iG}_{a,b,t} & \text{is distributed as} & p_\text{invGauss} \( \nu = a t , \lambda = \frac{a^3}{b^2}  t^2 \). \>
In order to sample the distribution of $m(t)$, which corresponds to the final sink particle masses, we only need a sample of inverse Gaussian variates $\text{iG}_{a,b,t}$.

Fig. \ref{figure_imf_fit} shows the distribution functions obtained from the semi-analytic modelling (red), comparing to the final mass distribution of the collapsing, accreting and strongly accreting sinks (black).
The left column of plots corresponds to stochastic growth where the fluctuations in the accretion rates  of the same magnitude as in the simulation, which is reduced by a factor of 10 in the right column.
We obtain $a$ and $b$ from the lognormal fit of the fluctuations in the accretion rates performed in Section \ref{subsection_accretion_individual}, which gave estimates of the parameters $\mu_\text{W}= -11.44$ and $\sigma_\text{W}= 0.74$.
This corresponds to an average $a=1.41$ and standard deviation $b=1.19$ in units where mass is in \Msun\ and time in $10^5$ yr.

The top panels assume only fluctuations in the accretion rates, no distribution of seed masses or growth times.
As seed mass we chose the average initial collapse mass $\bar{m_\text{coll}} = 0.23\ \Msun$ (blue line) and as growth time the average growth time $\bar{t} = 1.07 \ \times 10^5\ \text{yr}$.
Due to the strictly positive distribution of the accretion rates the modelled stochastic accretion does not allow for masses to be lower than the initial mass.
Thus fluctuations in the accretion rates alone are not sufficient to explain the mass function.
However, with the observed level of fluctuations a power-law tail appears at high masses, similar to the numerical simulation.
With a smaller level of fluctuations,  as in the top right panel, this is not the case.

Fluctuations in the accretion rates and a distribution of seed masses is considered in the middle panels of Fig. \ref{figure_imf_fit}.
We chose as seed masses the initial collapse masses from the numerical simulation, their distribution is shown a the blue histogram.
Growth time is again constant $=\bar{t}$ for the modelled sample.
For the observed level of fluctuations the agreement with the numerical simulation is better than without a distribution of seed masses, except at low masses, where the model under-predicts.
Again, lower levels of fluctuations in the accretion rates are not reproducing the numerical simulation.

All three factors, fluctuations in the accretion rates, a distribution of seed masses and a distribution of growth times is considered in the bottom panels of Fig. \ref{figure_imf_fit}.
Both the distribution of seed masses and the distribution of growth times are taken from the numerical simulation.
With all three factors as in the simulation the model and numerical distributions agree very well.
Only at very high masses the numerical simulation seems to have somewhat less sinks compared to the model.
With a lower level of fluctuations the model distribution covers a narrower mass range compared to the simulation.

Thus, we conclude that for the accreting sinks the shape of the final mass function is not dominated by one factor, but requires all three, fluctuations in the accretion rates, a distribution of seed masses and a distribution of growth times.
The final mass function of all sinks, which contains both only collapsing and accreting sinks, is also shaped by all three factors.
However, as there are about as many collapse-only as accreting sinks, the distribution of collapse masses will be more important in the lower mass range.
This is perhaps somewhat perceptible in Fig. \ref{figure_class_imf}, where final sink mass  function the mass range 0.1--1\ \Msun\ appears somewhat flatter than in the mass range 1--30\ \Msun.
The range 0.1--1\ \Msun\ is the overlap region between collapse-only and accreting sinks.
In the high-mass tail the distribution of accretion rates and growth times is more important.

\section{Summary}\label{sec_conclusions}

We have analysed a hydrodynamical sink-particle simulation with a barotropic equation of state for the distribution and mass dependence of the accretion rates, the distribution of growth times and the distribution of initial collapse masses.
We find that all these aspects are shaping the sink particle mass function. 
In detail we find:
\begin{enumerate}

\item  After an initial collapse phase sinks grow in episodes of accretion and can have long quiescent phases. During an episode the accretion rate follows shows a sharp rise followed by an exponential decay.

\item 
In about 50 per cent of sinks their mass is mainly set by an initial collapse phase while in 50 per cent of sinks acquire most of their mass through an extended accretion phase. 

\item The accretion rates follow $\dot{m} \propto m^{2/3}$ as expected from competitive accretion in a gas-dominated potential as predicted by \citet{BonnellBateClarke-2001a,BonnellClarkeBate-2001a}.

\item The fluctuations in the accretion rates follow a lognormal distribution, which is likely a consequence of the lognormality of a turbulent gas density.

\item The masses after the initial collapse follow roughly a lognormal distribution, with some evidence of power-law tails.

\item The growth times follow an exponential distribution but tapered at very long times.

\item The fluctuations in the accretion rates, the distribution of initial collapse masses and the distribution of growth times all shape the final sink mass function. The sink growth can be modelled as a non-linear stochastic process \cite[cf. ][]{Maschberger-2013b}.

\end{enumerate}

 We have thus shown that {\it in the simulations} the resulting sink mass function is not simply related to the accretion rate-mass relation, as proposed by \citet{Zinnecker-1982a} and \citet{BonnellBateClarke-2001a,BonnellClarkeBate-2001a}.
 {\it In particular a Salpeter-like upper IMF does not imply Bondi--Hoyle accretion}.

Finally we stress that of course there are many physical effects that are omitted or poorly modelled in the simulations and we might expect that protostellar feedback, magnetic fields or resolution effects could in principle affect the way that these three effects combine to form a resulting mass function. 
Although we can confidently assert that all three effects are important in the simulations we cannot {\it necessarily} claim that the observed IMF (which is broadly similar to the sink mass function) {\it must} also result from the same combination of these factors. 
Indeed, the sink mass function produced in the simulation can also be generated by a variety of other parameter choices (which are inconsistent with the simulation parameters). 
Thus, although we have presented a comprehensive analysis of the production of the mass function in the simulations, it remains the case that the observed IMF cannot on its own uniquely constrain the physics of star formation.

\section*{Acknowledgements}
We are grateful for the constructive report of the referee and thank Christophe Becker for inspiring discussions about the manuscript.
TM acknowledges funding via the ANR 2010 JCJC 0501 1 `DESC' (Dynamical Evolution of Stellar Clusters).
IAB acknowledges funding from the European Research Council for the FP7 ERC advanced grant project ECOGAL.
Some of this work was done as part of an International Team at the International Space Science Institute, Bern, Switzerland.

\bibliographystyle{mn2e}
\bibliography{refs_tm}

\label{lastpage}
\bsp

\end{document}